\documentclass[journal,12pt,onecolumn,draftclsnofoot,a4paper]{IEEEtran}

\usepackage[T1]{fontenc}
\usepackage{amsmath,amssymb,amsthm}
\usepackage{cite}
\usepackage{xcolor}
\usepackage{microtype}
\usepackage{comment}
\usepackage[acronym,shortcuts]{glossaries}
\usepackage{array}
\usepackage{bbold}
\newcommand{\bbzero}{\text{\usefont{U}{bbold}{m}{n}0}}
\newcommand{\bbone}{\text{\usefont{U}{bbold}{m}{n}1}}

\newtheorem{theorem}{Theorem}
\newtheorem{lemma}{Lemma}
\newtheorem{proposition}{Proposition}

\newtheorem{remark}{Remark}
\newtheorem{definition}{Definition}
\newtheorem{example}{Example}
\newtheorem*{restatedthm}{\restatedname}
\newcommand{\restatedname}{}
\newenvironment{restatement}[1]{\renewcommand{\restatedname}{#1}\begin{restatedthm}}{\end{restatedthm}}

\newacronym{GMR}{GMR}{Guruswami, Micciancio, and Regev}
\newacronym{LDPC}{LDPC}{low-density parity-check}

\newcommand{\F}{\mathbb{F}}
\newcommand{\wt}{\operatorname{wt}}
\newcommand{\CR}{\rho}
\newcommand{\nnz}{\operatorname{nnz}}
\newcommand{\PiP}{\Pi_2^{\mathrm P}}
\newcommand{\empt}{\varnothing}

\title{Hardness and Inapproximability of the Covering Radius for  LDPC Codes}

\author{Massimo Battaglioni and Fabrizio Marinelli%
\thanks{Massimo Battaglioni and Fabrizio Marinelli are with the Department of Information Engineering, Universit\`a Politecnica delle Marche, 60131 Ancona, Italy.}%
\thanks{Corresponding author: Massimo Battaglioni (e-mail: m.battaglioni@univpm.it).}}

\begin{document}

\maketitle
\glsresetall

\begin{abstract}
Given a binary parity-check matrix $H$ and an integer $r$, the covering-radius decision problem asks whether every attainable syndrome can be represented by an error vector of Hamming weight at most $r$. We study this problem for sparse parity-check matrices with bounded row and column weights. Using a bounded-occurrence quantified-SAT reduction, we prove that the problem is $\Pi_2^{\mathrm P}$-complete for matrices with maximum row and column weight three, and remains $\Pi_2^{\mathrm P}$-hard under a constant multiplicative gap. Starting from Vertex Cover on 3-regular graphs, we also obtain an explicit NP-hard approximation factor under the same weight constraints. We remark that weight three is the exact threshold for hardness: if either the maximum row weight or the maximum column weight is at most two, the covering radius can be computed in polynomial time.

\end{abstract}


\glsresetall

\section{Introduction}

The covering radius $\CR(C(H))$ of a binary linear code $C(H)$ is the largest Hamming distance between a vector in the ambient space and its nearest codeword~\cite{CohenHonkalaLitsynLobstein1997}. Given an integer $r\ge0$, deciding whether $\CR(C(H))\le r$ has a universal--existential form: every attainable syndrome must have a representative error vector of weight at most $r$.

In coding theory, sparsity is naturally defined for a family of parity-check matrices rather than for a single matrix. This raises the question of how sparsity restrictions on the parity-check matrices representing the considered codes affect the computational complexity of the covering-radius problem. We distinguish two notions based on restrictions that arise in the \ac{LDPC} literature\footnote{We impose no asymptotic minimum-distance or girth requirement on the families of \ac{LDPC} codes considered here~\cite{Gallager1962,Tanner1981,RichardsonUrbanke2008}.}: \emph{weak sparsity}, where either the maximum column weight or the maximum row weight of the parity-check matrix is uniformly bounded, and \emph{strong sparsity}, where both are uniformly bounded. Formal definitions are given in Section~\ref{sec:preli}. Accordingly, we refer to \ac{LDPC} codes satisfying weak or strong sparsity as \emph{weakly sparse} and \emph{strongly sparse} \ac{LDPC} codes, respectively. We show that the covering-radius problem is polynomial-time solvable when either the maximum row weight or the maximum column weight is at most two, whereas weight three already suffices for $\Pi_2^{\mathrm P}$-completeness. We then prove that this hardness persists when the row and column weights are simultaneously bounded by three.

A concrete motivation for this work comes from short \ac{LDPC}-coded telecommand systems~\cite{bluebook}. In this setting, the tail sequence is designed to be as far as possible from the set of codewords under a chosen metric~\cite{GiulianiAccess,BattaglioniTAES2025}. Under the Hamming metric, the maximum achievable distance of a tail sequence from the set of codewords is precisely the covering radius. However, the complexity results established in this paper characterize the computational difficulty of the problem over \emph{families} of \ac{LDPC} codes and do not imply that any fixed short code instance is computationally intractable.

\subsection{Related Work}

The covering radius is a classical parameter of coding theory, with an extensive literature but comparatively limited work on its computational complexity.
The survey~\cite{CohenKarpovskyMattsonSchatz1985} collects the principal bounds, exact values known at the time, and connections with other code parameters, while~\cite{CohenHonkalaLitsynLobstein1997} gives a systematic treatment of covering codes and covering radius. This literature provides much of the terminology and many of the structural tools surrounding the problem considered here, but it mainly studies bounds, exact values, and constructions rather than computational complexity.

Before the hardness of covering radius was studied directly, several fundamental coding problems had already been shown to be computationally hard.
Berlekamp, McEliece, and van Tilborg proved NP-completeness results for general decoding and related codeword-weight problems~\cite{BerlekampMcElieceVanTilborg1978}. Vardy later proved that computing the minimum distance of a linear code is NP-hard~\cite{Vardy1997}, and Dumer, Micciancio, and Sudan established strong inapproximability results for minimum distance~\cite{DumerMicciancioSudan2003}.
McGregor and Milenkovic further showed that approximating the minimum size of an $(a,b)$-trapping set in Tanner graphs is NP-hard, even for strongly sparse \ac{LDPC} codes~\cite{McGregorMilenkovic2010}.\footnote{The minimum-distance problem corresponds to the \(b=0\) special case, where minimizing \(a\) yields exactly the minimum distance of the code.} Jia et al.~\cite{JiaPengLiuWangYan2026} recently proved NP-completeness of the minimum distance problem for weakly sparse \ac{LDPC} codes whose parity-check matrices have any fixed column weight at least three. For regular, strongly sparse \ac{LDPC} codes, they proved NP-completeness when both the column and row weights are fixed and at least three. They also proved W[1]-completeness of an exact-weight parameterized variant for every fixed column weight at least three.

Covering radius, however, has a different quantifier structure: rather than asking for one low-weight nonzero codeword or one close codeword, it asks whether \emph{every} syndrome admits a low-weight representative. McLoughlin established $\Pi_2^{\mathrm P}$-completeness of the exact covering-radius decision problem for binary linear codes~\cite[Theorem 4]{McLoughlin1984}.  \ac{GMR} subsequently developed the approximation complexity of covering radius for linear codes~\cite{GuruswamiMicciancioRegev2005}. Their results show NP-hardness of approximation within any constant factor and $\Pi_2^{\mathrm P}$-hardness for a particular constant factor.  These results apply to general linear codes and do not address the covering-radius problem under sparsity constraints.

For cycle codes of graphs, the covering radius has a particularly simple characterization in terms of $T$-joins: it is the maximum, over all vertex sets $T$ of even cardinality, of the minimum size of a $T$-join~\cite{SoleZaslavsky1993}. Frank's results~\cite{Frank1993} imply that this quantity can be computed in polynomial time.

To the best of our knowledge, no previous work has characterized the computational complexity or approximability of the covering radius problem under fixed bounds on the row and column weights of the parity-check matrix. This motivates our paper.

\subsection{Main Results and Contributions}

Let $d_{\mathrm v},d_{\mathrm c}\geq 1$ be fixed integers denoting upper bounds on the maximum column and row weights, respectively, of a parity-check matrix.%
\footnote{The notation is based on the Tanner graph representation: the letter $d$ stands for degree, while the subscripts $\mathrm v$ and $\mathrm c$ refer to variable nodes and check nodes, respectively.} We denote by $\mathrm{CRP}(\cdot,d_{\mathrm c})$ and $\mathrm{CRP}(d_{\mathrm v},\cdot)$ the weakly sparse covering-radius problems, i.e., the problems under, respectively, only a row-weight bound and only a column-weight bound. We write $\mathrm{CRP}(d_{\mathrm v},d_{\mathrm c})$ for the strongly sparse version, in which both bounds are imposed.  For $\alpha>1$, $\mathrm{GapCRP}_{\alpha}(\cdot, \cdot)$ asks to distinguish covering radius at most $r$ from covering radius greater than $\alpha r$.%

\begin{theorem}[Bounded weights]\label{thm:bounded}
\leavevmode
\begin{enumerate}
\item[(i)] For every fixed $d\le2$, $\mathrm{CRP}(d,\cdot)$ and $\mathrm{CRP}(\cdot,d)$ are in P. Hence $\mathrm{CRP}(d_{\mathrm v},d_{\mathrm c})$ is in P whenever $d_{\mathrm v}\le2$ or $d_{\mathrm c}\le2$.
\item[(ii)] For all fixed $d_{\mathrm v},d_{\mathrm c}\ge3$, $\mathrm{CRP}(d_{\mathrm v},d_{\mathrm c})$, $\mathrm{CRP}(\cdot,d_{\mathrm c})$ and $\mathrm{CRP}(d_{\mathrm v},\cdot)$ are $\PiP$-complete.
\item[(iii)] There is a constant $c>0$ such that $\mathrm{GapCRP}_{1+c}(3,3)$ is $\PiP$-hard.
\end{enumerate}
\end{theorem}

Starting from the Vertex Cover Problem on 3-regular  graphs~\cite{ChlebikChlebikova2006}, our reductions also give an explicit approximation factor.

\begin{theorem}\label{thm:explicit}
$\mathrm{GapCRP}_{1882/1881}(3,3)$ is NP-hard.
\end{theorem}

By contrast, $\mathrm{GapCRP}_3(3,3)$ is in P by Proposition~\ref{prop:dv-approx}. A bound on the row weight alone also yields a constant factor: $\mathrm{GapCRP}_{5.43}(\cdot,3)$ is in P by Proposition~\ref{prop:dc-approx}.

For the problems of Theorems~\ref{thm:bounded} and~\ref{thm:explicit}, full row rank and pairwise distinct nonzero columns can be assumed without loss of generality (Section~\ref{subsec:normalization}). Moreover, the hardness reductions behind both theorems directly produce parity-check matrices with pairwise distinct nonzero columns. Hence every nonzero code they produce has minimum distance at least three, since a codeword of weight one or two would correspond to a zero column or to two equal columns.

The hardness proofs we propose begin with a source reduction that produces a parity-check matrix of column weight at most three, with a covering-radius threshold determined by the source instance. For $\Pi_2^{\mathrm P}$-hardness, we use a localized version of the $\forall\exists$-$3$-SAT construction of~\cite[Theorem~5.5]{GuruswamiMicciancioRegev2005}, applied to bounded-occurrence sources with perfect completeness~\cite{DockerDornLinzSemple2020,HavivRegevTaShma2007}. For explicit NP-hard factors, we use the Set-Cover construction of~\cite{GuruswamiMicciancioRegev2005} applied to Vertex Cover on 3-regular  graphs. We then repeatedly apply a \textit{cycle-based row-splitting} operator that reduces all row weights to three while preserving the weights of the pre-existing columns, at the cost of a controlled additive increase in the covering radius. This allows us to transfer both exact thresholds and constant multiplicative gaps.
Moreover, we introduce a \textit{contracted cycle-based row-splitting} operator, which removes one row from each split and thereby reduces the additive increase in the covering radius by one, leading to improved explicit approximation factors.

\subsection{Paper Organization}

The paper is organized as follows. Section~\ref{sec:preli} defines the problems, and Section~\ref{sec:tractable} proves Theorem~\ref{thm:bounded}(i). Section~\ref{sec:row-reduction} develops the row-splitting constructions and the tools for transferring exact thresholds and multiplicative gaps. Section~\ref{sec:pi2} proves $\Pi_2^{\mathrm P}$-completeness of the $\mathrm{CRP}$ and $\Pi_2^{\mathrm P}$-hardness of $\mathrm{GapCRP}$ for a constant approximation gap under bounded-weight restrictions. Section~\ref{sec:np} shows that, for column weight at most $d_{\mathrm v}$, the rank approximates the covering radius within a factor $d_{\mathrm v}$ in polynomial time, and within a constant factor depending only on $d_{\mathrm c}$ for row weight at most $d_{\mathrm c}$, and gives explicit NP-hard approximation factors using Vertex Cover on 3-regular graphs as source problem. Section~\ref{sec:concl} concludes the paper.

\section{Problem Formulation}\label{sec:preli}

We first introduce the notation for binary linear codes, formalize the notions of weak and strong sparsity, and define the exact and gap covering-radius problems considered throughout the paper. The source problems of our reductions are introduced where they are used, in Sections~\ref{sec:pi2} and~\ref{sec:np}.

\subsection{Notation}

We write $\F_2=\{0,1\}$ for the binary field. Vectors are regarded as row vectors. For a vector $x$, $x^{\mathsf T}$ denotes its transpose, and $\wt(x)$ its Hamming weight. For vectors $x,y$ of the same length,
\[
d_{\mathrm H}(x,y)=\wt(x-y)
\]
denotes their Hamming distance. All-zero and all-one vectors are referred to as $\bbzero$ and $\bbone$, respectively.

For a binary matrix $H\in\F_2^{s\times N}$, $H_{ij}$ denotes the entry in row $i$ and column $j$.
We write
\[
\nnz(H)=|\{(i,j):H_{ij}=1\}|
\]
for the number of nonzero entries of $H$, and use
$\operatorname{rank}(H)$ and $\operatorname{im}(H)$
for its rank and image over $\F_2$.

Let ${\bf 1}_S$ denote the incidence vector of an index set $S$, and $e_i$ the $i$th standard unit vector of the ambient space. Throughout the paper, a partition ${\cal P} = \{P_1,\ldots,P_k\}$ of a set $S$ is understood to have $k \ge 2$ non-empty parts.

\subsection{Linear Codes and Sparsity}

A binary linear code of length $N$ is a linear subspace $C\subseteq\F_2^N$. A matrix $H\in\F_2^{s\times N}$ is a parity-check matrix for $C$ if
\[
C=C(H)=\{c\in\F_2^N:Hc^{\mathsf T}= \bbzero\}.
\]
Rank-deficient matrices $H$ and repeated columns are generally permitted.  The code dimension is $N-\operatorname{rank}(H)\geq N-s$. The maximum column and row weights of the parity-check matrix are denoted by
\[
\Delta_{\mathrm v}(H)=\max_{1\le j\le N}\wt(He_j^{\mathsf T}),
\qquad
\Delta_{\mathrm c}(H)=\max_{1\le i\le s}\wt(e_iH),
\]
respectively.

\begin{definition}[Weak and strong sparsity]\label{def:sparsity}
Let $\{H_t\}$ be a family of binary parity-check matrices. The family is \emph{weakly sparse} if, for some constants $d_{\mathrm v},d_{\mathrm c}$ independent of $t$, at least one of the following conditions holds: the bounded-column-weight condition
$$
\Delta_{\mathrm v}(H_t)\le d_{\mathrm v}\qquad\forall t;
$$
or the bounded-row-weight condition
$$
\Delta_{\mathrm c}(H_t)\le d_{\mathrm c}\qquad\forall t.
$$ 
The family is \emph{strongly sparse} if both these conditions hold.
\end{definition}

\subsection{Problem Definitions}

For $q\in\operatorname{im}(H)$, define
\[
\lambda_H(q)=\min\{\wt(e):He^{\mathsf T}=q\}.
\]
Then, the covering radius of $C(H)$ is defined as
\begin{equation}\label{eq:syndrome-radius}
\CR(C(H))=\max_{q\in\operatorname{im}(H)}\lambda_H(q).
\end{equation}
If $H$ has full row rank, the maximum is over all $q\in\F_2^s$. The \emph{covering-radius problem} (CRP) asks for the computation of $\CR(C(H))$. In its decision version, given a nonnegative integer $r$, an instance $(H,r)$ is a YES instance if $\CR(C(H))\le r$ and a NO instance otherwise. Unless otherwise specified, CRP will refer to its decision version throughout what follows.

For fixed positive integers $d_{\mathrm v}$ and $d_{\mathrm c}$, let $\mathrm{CRP}(d_{\mathrm v},\cdot)$ and $\mathrm{CRP}(\cdot,d_{\mathrm c})$ denote the weakly sparse CRPs, obtained by restricting the parity-check matrices to those satisfying, respectively, the bounded-column-weight condition $\Delta_{\mathrm v}(H)\le d_{\mathrm v}$ and bounded-row-weight condition $\Delta_{\mathrm c}(H)\le d_{\mathrm c}$. 
Under simultaneous row- and column-weight bounds, let $\mathrm{CRP}(d_{\mathrm v},d_{\mathrm c})$ denote the strongly sparse CRP.

$\Pi_2^{\mathrm P}$ membership holds for all the above CRP variants, since checking $\CR(C(H))\le r$ is equivalent to checking whether, for every $q\in\F_2^s$, 
$$
q\in\operatorname{im}(H)
\Longrightarrow
\exists e\in\F_2^N:
\left(He^{\mathsf T}=q\land\wt(e)\le r\right);
$$
or, equivalently, that for every $q\in\F_2^s$
$$
\exists e\in\F_2^N:
\left[
q\notin\operatorname{im}(H)
\;\lor\;
\left(He^{\mathsf T}=q\land\wt(e)\le r\right)
\right],
$$
and this condition has the form of a $\Pi_2^{\mathrm P}$ predicate, since the $\operatorname{im}(H)$ membership  can be tested in polynomial time.

\subsection{Gap Versions}\label{subsec:gap-def}

We also consider the gap version $\mathrm{GapCRP}_{\alpha}$ of CRP, formulated as a promise problem. For $\alpha>1$, an instance $(H,r)$ of $\mathrm{GapCRP}_{\alpha}(\cdot,\cdot)$ is a YES instance if $\CR(C(H))\le r$ and a NO instance when $\CR(C(H))>\alpha r$. Instances satisfying
\[
r<\CR(C(H))\le \alpha r
\]
are outside the promise and may be answered arbitrarily.

Any algorithm for CRP
on a class of matrices solves $\mathrm{GapCRP}_{\alpha}(\cdot,\cdot)$ on promised inputs from the same class: promised YES instances satisfy $\CR(C(H))\le r$, while promised NO instances satisfy $\CR(C(H))>\alpha r\geq r$. Hence hardness of $\mathrm{GapCRP}_{\alpha}(\cdot,\cdot)$ on a class implies hardness of CRP on that class. Moreover, since CRP $\in\Pi_2^{\mathrm P}$ and $\alpha > 1$, the same $\Pi_2^{\mathrm P}$ predicate distinguishes the promised YES and NO
instances of $\mathrm{GapCRP}_{\alpha}$. Hence $\mathrm{GapCRP}_{\alpha}\in \Pi_2^{\mathrm P}$.

\subsection{Normalization}\label{subsec:normalization}

\begin{lemma}\label{lem:copies}
Let $H'$ be obtained from $H$ by appending $\bbzero$-columns
or copies of columns already present in $H$. Then $\lambda_{H'}(q)=\lambda_H(q)$ for every $q\in\operatorname{im}(H)$, and consequently $\CR(C(H'))=\CR(C(H))$.
\end{lemma}

\begin{IEEEproof}

The appended columns do not change the column space, so $\operatorname{im}(H')=\operatorname{im}(H)$. Since $h+h=\bbzero$ for every $h\in\F_2^s$, any minimum-weight representation of $q$ can be assumed to use distinct columns. Since every column of $H$ belongs to $H'$, and the additional columns of $H'$ are either $\bbzero$-columns or copies of columns of $H$, any minimum-weight representation of $q$ using columns of $H'$ can be turned, simply by removing $\bbzero$-columns and pairs of identical columns, into a representation of $q$ using distinct columns of $H$. Conversely, every representation using columns of $H$ is also available in $H'$. Hence,
$$
\lambda_{H'}(q)=\lambda_H(q) \qquad \forall q\in\operatorname{im}(H)
$$
and the radius identity follows from~\eqref{eq:syndrome-radius}. 
\end{IEEEproof}

\begin{lemma}[Normalization]\label{lem:normalization}
Let $H\in\F_2^{s\times N}$ have at least one nonzero column. In polynomial time one can construct a matrix $\widetilde H$ with full row rank and pairwise distinct nonzero columns such that
\[
\CR(C(\widetilde H))=\CR(C(H)),\qquad
\operatorname{rank}(\widetilde H)=\operatorname{rank}(H),
\]
\[
\Delta_{\mathrm v}(\widetilde H)\le\Delta_{\mathrm v}(H),\qquad
\Delta_{\mathrm c}(\widetilde H)\le\Delta_{\mathrm c}(H).
\]
\end{lemma}
\begin{IEEEproof}
Let $\widetilde H$ be the matrix obtained from $H$ by removing all zero columns and all but one copy of each repeated column. By Lemma~\ref{lem:copies}, $C(H)$ and $C(\widetilde H)$ have the same covering radius; the column space, and hence the rank, is unchanged as well. Next, keep in $\widetilde H$ only a set $B$ of rows forming a basis of the row space. This does not change the code, and hence its covering radius. Every deleted row is a linear combination of the rows in $B$. Therefore, the entries of any column on the deleted rows are completely determined by its entries on the rows in $B$. Consequently, two columns that become equal after the row deletion must already have been equal before it, and a column that becomes $\bbzero$ must already have been $\bbzero$. Hence distinct nonzero columns remain distinct and nonzero. Finally, deleting columns cannot increase a row weight, and deleting rows cannot increase a column weight. Both reductions can be performed in polynomial time.
\end{IEEEproof}

Consequently, $\mathrm{CRP}(d_{\mathrm v},\cdot)$, $\mathrm{CRP}(\cdot,d_{\mathrm c})$, $\mathrm{CRP}(d_{\mathrm v},d_{\mathrm c})$ and their gap versions may be restricted to full-row-rank parity-check matrices with pairwise distinct nonzero columns. Indeed, if $H$ is the zero matrix, then $C(H)=\F_2^N$ and $\CR(C(H))=0$, so $(H,r)$ is a YES instance for every $r\ge 0$, for both the decision and the gap versions, and the reduction outputs the fixed YES instance $([1],1)$: the matrix $[1]$ has full row rank, a single nonzero column, row and column weight one, and $\CR(C([1]))=1$, since $C([1])=\{0\}$.

\begin{definition}[3-systematic matrix]\label{def:admissible}
A parity-check matrix $H\in\F_2^{s\times N}$ is called \emph{3-systematic}\footnote{We use the term \emph{systematic} up to column permutations, since the identity submatrix need not occupy predetermined positions.} if (i) every unit vector $e_i$, $1\le i\le s$, occurs as a column of $H$, and 
(ii) its columns are nonzero, pairwise distinct, and of weight at most three.\end{definition}

Every 3-systematic matrix has full row rank, so $\operatorname{im}(H)=\F_2^s$ and
\[
\CR(C(H))=\max_{q\in\F_2^s}\lambda_H(q).
\]
Both matrices constructed in Sections~\ref{sec:pi2} and~\ref{sec:np} satisfy the conditions of Definition~\ref{def:admissible}.

\section{Polynomial-Time Cases}\label{sec:tractable}

We first establish the polynomial-time side of the classification. If the maximum row weight is at most two, the code decomposes into elementary component codes, and its covering radius can be computed in polynomial time. If the maximum column weight is at most two, the code is a cycle code, for which the covering radius is also polynomial-time computable.

\subsection{Maximum Row Weight Two}

Let $H\in\F_2^{s\times N}$ be a parity-check matrix.

\begin{proposition}\label{prop:row-two}
If $\Delta_{\mathrm{c}}(H)\leq 2$, then $\CR(C(H))$ is computable in polynomial time.
\end{proposition}

\begin{IEEEproof}
Rows equal to $\bbzero$ impose no constraint and may be discarded. A row of weight one, with support $\{i\}$, reads $x_i=0$, whereas a row of weight two, with support $\{i,j\}$, reads $x_i+x_j=0$, i.e.\ $x_i=x_j$.
Let $F=(V,\mathcal E)$ be the graph with $V=\{1,\dots,N\}$ the set of code coordinates and $\mathcal E$ the set of edges corresponding to the weight-two rows. A vertex is \emph{marked} if its coordinate occurs in a weight-one row. Then
\[
  C(H)=\bigl\{x\in\mathbb F_2^N : x_i=0 \text{ for every marked } i,\ \ x_i=x_j \text{ for every }\{i,j\}\in \mathcal E\bigr\}.
\]
Along any path in $F$ the coordinates coincide, so all coordinates in a connected component are equal.
Since the connected components of $F$ partition the coordinates into disjoint sets, $C(H)$ is the direct product of its restrictions to the components. The Hamming distance is additive over disjoint blocks of coordinates, so $\CR(C(H))$ is the sum of the covering radii of the component codes.
Let $Q$ be a component with $\ell$ vertices.
If $Q$ contains a marked vertex, all its coordinates are forced to zero: the only codeword on $Q$ is $\bbzero\in\mathbb F_2^\ell$, hence its covering radius is $\ell$.
If $Q$ is unmarked, its component code is the length-$\ell$ repetition code, with covering radius $\lfloor \ell/2\rfloor$.
Therefore
\[
  \CR(C(H))=\sum_{Q\ \text{marked}}|Q|+\sum_{Q\ \text{unmarked}}\lfloor |Q|/2\rfloor .
\]
The components and their marks can be found by a graph traversal in time $O(|V|+|\mathcal E|)$. Hence $\CR(C(H))$ is computable in polynomial time.
\end{IEEEproof}

\subsection{Maximum Column Weight Two}

Let $F=(V,\mathcal E)$ be a graph and let $H_{F}\in\F_2^{V\times \mathcal E}$ be its vertex--edge incidence matrix. The \emph{cycle code} of $F$ is $C(H_{F})=\ker H_{F}$, i.e., the set of edge subsets forming even-degree subgraphs. For $T\subseteq V$ of even cardinality, a \emph{$T$-join} is a set of edges whose odd-degree vertices are exactly the vertices of $T$. A graph is $2$-edge-connected if it is connected and has no bridge, i.e., no cut of cardinality one. An \emph{ear decomposition} of a $2$-edge-connected graph $F$ builds $F$ from a single vertex by successively adding \emph{ears}, namely paths (or circuits) whose endpoints already belong to the current graph and whose internal vertices are new. Let $\psi(F)$ be the minimum number of even-length ears over all ear decompositions of $F$.

\begin{lemma}[{Sol\'e--Zaslavsky~\cite[Theorem~3.2]{SoleZaslavsky1993}, Frank~\cite[Theorem~4.5]{Frank1993}}]
\label{lem:cycle-cr}
If $F=(V,\mathcal E)$ is $2$-edge-connected, then
\[
  \CR(C(H_{F}))=\frac{|V|-1+\psi(F)}{2},
\]
and this quantity can be computed in polynomial time.
\end{lemma}

\noindent
Indeed, $\CR(C(H_{F}))$ is the maximum, over even-size $T\subseteq V$, of the minimum size of a $T$-join in $F$ \cite{SoleZaslavsky1993}, and for $2$-edge-connected $F$ this maximum equals $(|V|-1+\psi(F))/2$ and an optimal ear decomposition can be found in polynomial time \cite{Frank1993}.

\begin{proposition}\label{prop:column-two}
If every column of $H$ has weight at most two, then $\CR(C(H))$ can be computed in polynomial time.
\end{proposition}

\begin{IEEEproof}
By Lemma~\ref{lem:copies}, we may assume that the columns of $H$ are nonzero and pairwise distinct.
Let $F=(V,\mathcal E)$ with $V=\{0,1,\ldots,s\}$, where a weight-two column with ones in rows $i,j$ becomes an edge $ij$, and a unit column in row $i$ becomes an edge $0i$. Then $H$ is $H_{F}$ with the row of vertex $0$ deleted.
Since the sum of all rows of $H_{F}$ within a connected component is zero over $\F_2$, the row of vertex $0$ is a combination of the other rows of its component, and deleting it does not change the kernel. Therefore $C(H)=C(H_{F})$.

The cycle code decomposes over the connected components of $F$, so their covering radii add. A bridge lies in no cycle, so its coordinate vanishes on every codeword and contributes exactly one to the covering radius. Removing the bridges leaves $2$-edge-connected components (isolated vertices contribute zero), whose covering radii are given by Lemma~\ref{lem:cycle-cr}. Components, bridges, and $\psi$ are computable in polynomial time, hence so is $\CR(C(H))$.
\end{IEEEproof}

\begin{restatement}{Theorem~\ref{thm:bounded}(i)}
For every fixed $d\le2$, $\mathrm{CRP}(d,\cdot)$ and $\mathrm{CRP}(\cdot,d)$ are in P. Hence $\mathrm{CRP}(d_{\mathrm v},d_{\mathrm c})$ is in P whenever $d_{\mathrm v}\le2$ or $d_{\mathrm c}\le2$.
\end{restatement}
\begin{IEEEproof}
Proposition~\ref{prop:column-two} gives a polynomial-time algorithm for $\mathrm{CRP}(d_{\mathrm v},\cdot)$ whenever $d_{\mathrm v}\le2$, while Proposition~\ref{prop:row-two} gives one for $\mathrm{CRP}(\cdot,d_{\mathrm c})$ whenever $d_{\mathrm c}\le2$. Since the additional row- or column-weight bound defining $\mathrm{CRP}(d_{\mathrm v},d_{\mathrm c})$ can be checked in polynomial time, the strongly sparse problem is also in $\mathrm P$ whenever $d_{\mathrm v}\le2$ or $d_{\mathrm c}\le2$.
\end{IEEEproof}

\section{Row-Weight Reduction with Exact Radius Shifts}\label{sec:row-reduction}

We now focus on parity-check matrices with column weight at most three and aim to reduce the row weight to the same bound. As anticipated in the introduction, the source problems already provide matrices with column weight at most three. We iteratively apply cycle-based row-splitting operators to reduce the maximum row weight to any prescribed bound $D\ge3$ while keeping column weights at most three, with a precisely determined additive change in the covering radius. With $D=3$, this yields the bounded $(3,3)$ instances of Sections~\ref{sec:pi2} and~\ref{sec:np}.

The row-splitting constructions below add new coordinates to the original code. To analyze their effect on the covering radius, consider a binary linear code $C\subseteq\F_2^N$ and the following extension. Let  $\mathcal D\subseteq\F_2^\ell$ be a binary linear code, and let $\varphi:C\to\F_2^\ell$ be a linear map. 
The set of pairs $(x,z)$, each obtained by appending  a vector $z$ of the form
$$
z=w+\varphi(x), \qquad w\in \mathcal D,
$$
to a codeword $x\in C$, constitutes an extended code
$$
C' = \{(x,z):x\in C,\ z\in \mathcal D+\varphi(x)\}.
$$
For each $x \in C$, the set of possible values of $z$ is thus the coset $\mathcal D+\varphi(x)$ of the linear code $\mathcal D$, which we call the \emph{fiber} associated with $x$, with $\varphi(x)$ as a coset representative. In particular, the fiber associated with $x=0$ is $\mathcal D$, since $\varphi(0)=0$.

The following lemma gives a sufficient condition for this extension to increase the covering radius by exactly $\CR(\mathcal D)$.

\begin{lemma}[Fiber extension]\label{lem:fiber-extension}
Let $C\subseteq\F_2^N$ and $\mathcal D\subseteq\F_2^\ell$ be binary linear codes, and let $\varphi:C\to\F_2^\ell$ be a linear map.  Define the extended code
\[
C'=\{(x,z):x\in C,\ z\in \mathcal D+\varphi(x)\}
\]
and let $\mathcal K\subseteq\F_2^\ell$ be a binary linear code such that
\[
\mathcal D+\operatorname{im}(\varphi)\subseteq \mathcal K.
\]
If
\[
\CR(\mathcal D)=\CR(\mathcal K)=:\rho_0,
\]
then
\[
\CR(C')=\CR(C)+\rho_0.
\]
\end{lemma}

\begin{IEEEproof}
We first prove that
\[
\CR(C')\le \CR(C)+\rho_0.
\]
Consider any $(y,z)\in\F_2^{N+\ell}$. By the definition of covering radius, there exists $x\in C$ such that
\[
d_{\mathrm H}(y,x)\le \CR(C).
\]
For this $x$, every vector $(x,w + \varphi(x))$ with $w\in \mathcal D$ is a codeword of $C'$. Since $\CR(\mathcal D)=\rho_0$, we can choose $w\in \mathcal D$ such that
$d_{\mathrm H}(z-\varphi(x),w)\le \rho_0$ or, equivalently,
\[
d_{\mathrm H}(z,w+\varphi(x))\le \rho_0.
\]
By the additivity of the Hamming distance under concatenation, 
$$
d_{\mathrm H}\bigl((y,z),(x,w+\varphi(x))\bigr)
\le \CR(C)+\rho_0.
$$
It follows that
\[
\min_{c'\in C'} d_{\mathrm H}\bigl((y,z),c'\bigr)
\le \CR(C)+\rho_0.
\]
Since $(y,z)$ was arbitrary, taking the maximum over $(y,z)\in\F_2^{N+\ell}$ gives
$$
\CR(C')\le \CR(C)+\rho_0.
$$

We now prove the reverse inequality. By the definition of covering radius, there exist $y^\star\in\F_2^N$ and $z^\star\in\F_2^\ell$ such that $d_{\mathrm H}(y^\star,x)\ge\CR(C)$ for every $x\in C$ and $d_{\mathrm H}(z^\star,u)\ge \rho_0$ for every $u\in \mathcal K$. Let $(x,w+\varphi(x))$, with $x\in C$ and $w\in \mathcal D$, be an arbitrary codeword of $C'$. Since $w+\varphi(x)\in \mathcal D+\operatorname{im}(\varphi)\subseteq \mathcal K$, and $z^\star$ is at distance at least $\rho_0$ from every
element of $\mathcal K$, we have
\[
d_{\mathrm H}(z^\star,w+\varphi(x)) \ge \rho_0. 
\]
Hence, by the additivity of the Hamming distance under concatenation,
\[
\begin{aligned}
d_{\mathrm H}\bigl((y^\star,z^\star),(x,w+\varphi(x))\bigr)
&= d_{\mathrm H}(y^\star,x)
 + d_{\mathrm H}(z^\star,w+\varphi(x)) \\
&\ge \CR(C)+\rho_0 .
\end{aligned}
\]
Since $(x,w+\varphi(x))$ was an arbitrary codeword of $C'$, it follows that
\[
\min_{c'\in C'}d_{\mathrm H}\bigl((y^\star,z^\star),c'\bigr)
\ge \CR(C)+\rho_0.
\]
Taking the maximum over all vectors in $\F_2^{N+\ell}$ therefore gives
\[
\CR(C')\ge\CR(C)+\rho_0.
\]
Combining the two inequalities proves the claim.

\end{IEEEproof}

\subsection{Cycle Row Split}

We first introduce a cycle-based row-splitting operation that preserves the weight of each column of the original parity-check matrix $H$ and changes the covering radius $\CR(C(H))$ by an exactly known amount. We then apply it
to reduce the maximum row weight to any fixed bound $D\ge3$.

\begin{definition}[Cycle row split]\label{def:cycle-row-split}
Let $H\in\F_2^{s\times N}$, let $S_i$ be the support of row $i$ of $H$, and let $\mathcal P=\{P_1,\ldots,P_k\}$ be a partition of $S_i$ into $k\ge2$ parts. 
The \emph{cycle row split} \textsc{CRSplit}$(H, i, \mathcal P)$ of row $i$ with respect to $\mathcal P$ is the following operation. Introduce $2k$ new coordinates $z_1,\ldots,z_{2k}$, and replace row $i$ by $2k$ rows $W_1,R_1,\ldots,W_k,R_k$. In the original $N$ columns, each row $R_j$ is ${\bf 1}_{P_j}$ and each row $W_j$ is $\bbzero$. Let $G \in \mathbb F_2^{2k \times 2k}$ be the vertex-edge incidence matrix of the cycle graph
\[
R_k-W_1-R_1-W_2-R_2-\cdots-W_k-R_k,
\]
whose vertices are the rows $W_1,R_1,\ldots,W_k,R_k$ and whose $t$-th edge along the cycle is the coordinate $z_t$. Thus, $z_{2j}$ joins $W_j$ and $R_j$ for $1\le j\le k$, $z_{2j+1}$ joins $R_j$ and
$W_{j+1}$ for $1\le j\le k-1$, and $z_1$ joins $R_k$ and $W_1$, closing the cycle. On the new $2k$ columns, each row $W_j$ or $R_j$ equals the row of $G$ indexed by the same vertex, and all other rows of $H$ are extended by zeros. We denote the resulting matrix by $H'$.
\end{definition}

It is easy to see that every column of $H'$ inherited from $H$ has the same weight as in $H$, every new column has weight two, the rows $R_j$ have weight $|P_j|+2$, and the rows $W_j$ have weight two.

\begin{example}\label{ex:cycle-row-split}
Consider the parity-check matrix
\[
H=\begin{bmatrix}1&1&1&1&1&1&1\end{bmatrix}.
\]
Choose $P_1=\{1,2,3\}$ and $P_2=\{4,5,6,7\}$, so that $k=2$. Applying $\textsc{CRSplit}(H,1,\mathcal P)$ introduces four new coordinates $z_1,z_2,z_3,z_4$ and gives
\[
H'=
\begin{array}{c@{\;}c}
\begin{array}{r}
W_1\\
R_1\\
W_2\\
R_2
\end{array}
&
\left[
\begin{array}{ccccccc|cccc}
0&0&0&0&0&0&0&1&1&0&0\\
1&1&1&0&0&0&0&0&1&1&0\\
0&0&0&0&0&0&0&0&0&1&1\\
0&0&0&1&1&1&1&1&0&0&1
\end{array}
\right]
\end{array}.
\]
The right-hand block is the vertex-edge incidence matrix $G$ of the cycle graph $R_2-W_1-R_1-W_2-R_2$. The original columns retain weight one, each new column has weight two, the rows $R_1$ and $R_2$ have weights $|P_1|+2=5$ and $|P_2|+2=6$, and the maximum row weight decreases from seven to six.
\end{example}

\begin{lemma}\label{lem:cycle-row-splitting}
Let $H'$ be obtained  by applying \textsc{CRSplit}$(H, i, \mathcal P)$ to $H$, where $\mathcal P = \{P_1,\ldots,P_k\}$ is a partition of the support $S_i$ of  row $i$ of $H$.
Then,
\[
\CR(C(H'))=\CR(C(H))+k.
\]
Moreover, if the columns of $H$ are pairwise distinct and nonzero, then so are the columns of $H'$.
\end{lemma}

\begin{IEEEproof}

Consider the system $H'(x,z)^{\mathsf T}=\bbzero$, where $x\in\F_2^N$ collects the original coordinates and $z\in\F_2^{2k}$ the new ones.
Every row of $H$ not affected by the \textsc{CRSplit} operation is extended by zeros on $z$, so the corresponding row of $H'$ imposes the same condition on $x$. Let $\bar H\in\F_2^{2k\times N}$ be the matrix whose rows, ordered as $W_1,R_1,\ldots,W_k,R_k$, are $\bbzero,{\bf 1}_{P_1},\ldots,\bbzero,{\bf 1}_{P_k}$. 
Hence, $(x,z)\in C(H')$ if and only if $x$ satisfies all the rows of $H$ other than row $i$ and $(x,z)$ solves the subsystem 
\[
{\bar H}x^{\mathsf T} + {G}z^{\mathsf T} = \bbzero,
\] 
corresponding to the rows $W_j$ and $R_j$.
Let 
\[
\theta_j(x)=\sum_{\ell\in P_j}x_\ell, \qquad 1\le j\le k.
\]
Since every column of $G$ contains exactly two ones, summing all the equations of the subsystem eliminates the variables in $z$ and yields  
\[
\sum_{j=1}^k\theta_j(x)=0,
\]
which is therefore a necessary condition for the subsystem to admit a solution. 
Conversely, if $\sum_{j=1}^k\theta_j(x)=0$, then $\bar Hx^{\mathsf T}$ belongs to $\operatorname{im}(G)$. Indeed, since $G$ is the incidence matrix of a cycle of length $2k$, it has rank $2k-1$. The argument above shows that the entries of $Gz^{\mathsf T}$ sum to zero for every $z\in\F_2^{2k}$, i.e.,  $\operatorname{im}(G)$ is contained in the set of vectors of $\F_2^{2k}$ whose entries sum to zero; both have dimension $2k-1$, so they coincide. The entries of $\bar Hx^{\mathsf T}$ sum to $\sum_{j=1}^k\theta_j(x)=0$, hence  there exists $z\in\F_2^{2k}$ such that
\[Gz^{\mathsf T}=\bar Hx^{\mathsf T}.
\]

Since the sets $P_j$ partition $S_i$, the condition $\sum_{j=1}^k\theta_j(x)=0$ is equivalent to
\[
\sum_{\ell\in S_i}x_\ell=0.
\]
Thus the subsystem ${\bar H}x^{\mathsf T} + {G}z^{\mathsf T} = \bbzero$
is solvable if and only if $x$ satisfies the original row-$i$ parity check. Together with the rows not affected by the \textsc{CRSplit} operation,
this shows that the projection of $C(H')$ onto the original coordinates is exactly $C(H)$. 

Moreover, every row of $G$ also contains exactly two ones, so $G\bbone^{\mathsf T}=\bbzero$. Since $G$ is a $2k\times 2k$ matrix of rank $2k-1$, its kernel has dimension one, and therefore \[\ker(G)=\{\bbzero,\bbone\}\subseteq \mathbb F^{2k}_2.\] Hence, for a fixed $x\in C(H)$, if $z^\star$ is a solution of $Gz^{\mathsf T}=\bar Hx^{\mathsf T}$, the system has exactly two solutions, $z^\star$ and $z^\star+\bbone$. The two solutions differ in every coordinate, so fixing the value of any single coordinate of $z$ uniquely determines the solution. 

Let
\[
\mathcal D=C(G)=\{\bbzero,\bbone\}\subseteq \mathbb F^{2k}_2
\]
be the code with parity-check matrix $G$, i.e., the length-$2k$ repetition code.

Since the rows $W_j$ vanish on the original coordinates, the equation associated with
$W_j$ is
\[
z_{2j-1}+z_{2j}=0, \qquad 1\le j\le k,
\]
so every solution $z$ belongs to
\[
\mathcal K=\{z\in\F_2^{2k}:\ z_{2j-1}=z_{2j},\ 1\le j\le k\}.
\]
For each $x\in C(H)$, let $\varphi(x)$ be the unique solution with $z_1=0$. If $x,x'\in C(H)$, then $\varphi(x)+\varphi(x')$ is a solution for $x+x'$ whose first coordinate is zero, so it coincides with $\varphi(x+x')$. Hence 
$\varphi:C(H)\to \mathcal K$ is linear. The set of all extensions of a fixed
$x$ is therefore exactly
\[
\mathcal D+\varphi(x).
\]
Moreover, since $\mathcal D\subseteq \mathcal K$ and $\mathcal K$ is linear,
\[
\mathcal D+\operatorname{im}(\varphi)\subseteq \mathcal K.
\]

The repetition code $\mathcal D$ has covering radius $k$. The code $\mathcal K$ is the direct sum of $k$ copies of
the length-two repetition code $\{00,11\}$, which has covering radius
one. Hence
\[
\CR(\mathcal K)=k.
\]

The codes $C(H), C(H'), \mathcal D, \mathcal K$ satisfy the \textit{fiber extension} hypothesis (see Lemma~\ref{lem:fiber-extension}). Therefore,
\[
\CR(C(H'))=\CR(C(H))+k.
\]

It remains to prove that pairwise distinct columns are preserved. Assume that the columns of $H$ are pairwise distinct. For each inherited column, summing its entries in $R_1,\ldots,R_k$ recovers its original entry in row $i$, while all other original entries are unchanged. Hence inherited columns remain pairwise distinct and nonzero. Every inherited column vanishes on all rows $W_j$, whereas every new column has a one in one of these rows, so the two types cannot coincide.
The new columns are nonzero and pairwise distinct because they are the edge-incidence columns of a simple cycle of length $2k\ge4$.

\end{IEEEproof}

For a matrix $H$ and an integer $D\ge3$, define
\[
\Omega_{D}(H)=\sum_{i:\,\wt(e_iH)>D}\left\lceil\frac{\wt(e_iH)}{D-2}\right\rceil.
\]
For $D=3$, this is the total weight of the rows of $H$ of weight at least four. 

\begin{theorem}[Row-weight reduction]\label{thm:degree-three}Let $D\ge3$ be fixed. From any $H\in\F_2^{s\times N}$ 
one can construct in polynomial time a matrix $H^{(D)}$ such that
\[
\Delta_{\mathrm c}(H^{(D)})\le D,
\qquad
\Delta_{\mathrm v}(H^{(D)})\le\max\{\Delta_{\mathrm v}(H),2\},
\]
and
\[
\CR(C(H^{(D)}))=\CR(C(H))+\Omega_{D}(H).
\]
Pairwise distinct nonzero columns are preserved.
\end{theorem}

\begin{IEEEproof}
    For each row $i$ of $H$ of weight $d>D$, apply \mbox{\textsc{CRSplit}} to that row with $k=\lceil d/(D-2)\rceil$ and
each part $P_j$ containing 
at most $D-2$ elements of $S_i$; this is possible because $d > D-2 \ge 1$
gives $2\le k\le d$. The rows $R_j$ produced by the split then have weight at most $D$, while the rows $W_j$ have weight two. All other rows are only extended by zeros, so a split leaves the weights of the other rows unchanged, and the rows it creates have weight at most $D$
and are never split again. 
Applying the construction to each original row of weight greater than $D$ therefore gives $\Delta_{\mathrm c}(H^{(D)})\le D$. Each split leaves the weights of the pre-existing columns unchanged and introduces only columns of weight two, so
\[
\Delta_{\mathrm v}(H^{(D)})\le\max\{\Delta_{\mathrm v}(H),2\}.
\]
By Lemma~\ref{lem:cycle-row-splitting}, the split of a row of weight $d$ increases the covering radius by exactly $k$, and hence the total shift is $\Omega_{D}(H)$. Pairwise distinct nonzero columns are preserved by Lemma~\ref{lem:cycle-row-splitting}. Finally, $\Omega_{D}(H)\le\nnz(H)$, since $k\le d$, and splitting a row of weight $d$ into $k$ parts adds $2k\le 2d$ columns and $2k-1$ rows; so the construction is polynomial-time in the size of $H$.%
\end{IEEEproof}

\subsection{Contracted Cycle Row Split}

By Lemma~\ref{lem:cycle-row-splitting}, splitting a row into $k$ parts shifts the covering radius by $k$, and Theorem~\ref{thm:degree-three} pays this shift for every split row. When a multiplicative gap is transferred, the resulting constant degrades with the ratio between the total shift and the threshold (Lemma~\ref{lem:gap-transfer} below). We now show that one row of each split can be removed, which lowers the shift of a single row split to $k-1$. Section~\ref{sec:np} uses this to improve the explicit constant of Theorem~\ref{thm:explicit}.

\begin{definition}[Contracted cycle row split]\label{def:contracted-cycle-row-split}
Let $H$, $i$ and $\mathcal P=\{P_1,\ldots,P_k\}$ be as in Definition~\ref{def:cycle-row-split}. The \emph{contracted cycle row split} \textsc{CCRSplit}$(H,i,\mathcal P)$ of row $i$ with respect to $\mathcal P$ is defined as \textsc{CRSplit}$(H,i,\mathcal P)$, except that the row $W_k$ is omitted, only $2k-1$ new coordinates $z_1,\ldots,z_{2k-1}$ are introduced, and $G$ is replaced by the vertex-edge incidence matrix $G''\in\F_2^{(2k-1)\times(2k-1)}$ of the cycle graph
\[
R_k-W_1-R_1-W_2-R_2-\cdots-W_{k-1}-R_{k-1}-R_k,
\]
whose $t$-th edge along the cycle is the coordinate $z_t$. We denote the resulting matrix by $H''$.
\end{definition}

\begin{example}\label{ex:contracted-cycle-row-split}
For $H$ and $\mathcal P$ as in Example~\ref{ex:cycle-row-split}, \textsc{CCRSplit}$(H,1,\mathcal P)$ introduces three new coordinates $z_1,z_2,z_3$ and gives
\[
H''=
\begin{array}{c@{\;}c}
\begin{array}{r}
W_1\\
R_1\\
R_2
\end{array}
&
\left[
\begin{array}{ccccccc|ccc}
0&0&0&0&0&0&0&1&1&0\\
1&1&1&0&0&0&0&0&1&1\\
0&0&0&1&1&1&1&1&0&1
\end{array}
\right]
\end{array}.
\]
The right-hand block is the vertex-edge incidence matrix $G''$ of the cycle graph $R_2-W_1-R_1-R_2$. The column $z_3$ of $H''$ is the sum of the columns $z_3$ and $z_4$ of the matrix $H'$ of Example~\ref{ex:cycle-row-split}, and the row $W_2$, being $\bbzero$, is omitted.
\end{example}

\begin{lemma}[Contraction]\label{lem:contraction}
The conclusions of Lemma~\ref{lem:cycle-row-splitting} also hold when \textsc{CRSplit} is replaced by \textsc{CCRSplit} and $H'$ by $H''$, except that
\[
\CR(C(H''))=\CR(C(H))+k-1.
\]
\end{lemma}
\begin{IEEEproof}
We adapt the proof of Lemma~\ref{lem:cycle-row-splitting}, replacing $G$ with $G''$. Up to the definition of $\mathcal K$, the argument relies only on the fact that $G$ is square, has exactly two ones in each row and column, and has rank one less than its order. These properties are also satisfied by $G''$, whose rank is $2k-2$. Hence the projection of $C(H'')$ onto the original coordinates is $C(H)$. Let
\[
\mathcal K=\{z\in\F_2^{2k-1}:\ z_{2j-1}=z_{2j},\ 1\le j\le k-1\}
\]
be the direct sum of $k-1$ copies of the length-two repetition code $\{00,11\}$ and of $\F_2$ on the coordinate $z_{2k-1}$, and let
\[
\mathcal D=C(G'')=\{\bbzero,\bbone\}\subseteq\F_2^{2k-1}
\]
be the length-$(2k-1)$ repetition code. For each $x\in C(H)$, let $\varphi(x)$ denote the unique extension of $x$ to the new coordinates satisfying $z_1=0$. Following the argument of Lemma~\ref{lem:cycle-row-splitting}, the set of all extensions of $x$ then forms the fiber $\mathcal D+\varphi(x)$, and $\mathcal D+\operatorname{im}(\varphi)\subseteq\mathcal K$. Since
\[
\CR(\mathcal D)=\left\lfloor\frac{2k-1}{2}\right\rfloor=k-1
\qquad\text{and}\qquad
\CR(\mathcal K)=k-1,
\]
the radius identity follows from Lemma~\ref{lem:fiber-extension}. Finally, the fact that $z_{2k-1}$ does not meet any row $W_j$ does not compromise its distinctness from the inherited columns, since it has ones in both rows $R_{k-1}$ and $R_k$, whereas any inherited column has at most one nonzero entry among the rows $R_j$, as the parts $P_j$ are disjoint. Moreover, the new columns remain pairwise distinct, as they correspond to the edges of a simple cycle of length $2k-1\ge3$.
\end{IEEEproof}

Omitting two or more of the rows $W_j$, to reduce the shift further, would violate the hypothesis of Lemma~\ref{lem:fiber-extension}. Indeed, if $a\ge2$ rows $W_j$ are omitted, $\mathcal D$ becomes the repetition code of length $2k-a$, while the remaining $k-a$ rows $W_j$ yield $\CR(\mathcal K)=k-a$. Since
\[
\CR(\mathcal D)=\left\lfloor\frac{2k-a}{2}\right\rfloor>k-a=\CR(\mathcal K),
\]
the required equality $\CR(\mathcal D)=\CR(\mathcal K)$ does not hold.

\subsection{Transfer of Thresholds and Relative Gaps}

The matrix transformations proposed in this section change the covering radius by an exactly determined additive amount $\sigma$, that is, $\CR(C(H'))=\CR(C(H))+\sigma$. Therefore,
\[
\CR(C(H))\le r
\quad\Longleftrightarrow\quad
\CR(C(H'))\le r+\sigma
\]
for any threshold $r$.

For a threshold value $r>0$, we define the relative gap of $\CR(C(H))$ with respect to $r$ as
\[
\gamma_r(H):=\frac{\CR(C(H))-r}{r}.
\]
The following lemma shows that a nonnegative relative gap is transferred from $H$ to $H'$, possibly with a reduction by a constant factor, provided that the shift $\sigma$ of the covering radius is bounded by a constant multiple of $r$.

\begin{lemma}[Relative gap transfer]\label{lem:gap-transfer}
Let $H$ and $H'$ satisfy $\CR(C(H'))=\CR(C(H))+\sigma$, 
and let $r>0$ and $\beta\ge0$ be such that $0\le\sigma\le\beta r$. If $\gamma_r(H)\ge0$, then
\[
\gamma_{r+\sigma}(H')\ge\frac{1}{1+\beta}\,\gamma_r(H).
\]
In particular, for every $\delta\ge0$, $\gamma_r(H)>\delta$ implies $\gamma_{r+\sigma}(H')>\delta/(1+\beta)$, and the same holds with $\ge$ in place of $>$.
\end{lemma}

\begin{IEEEproof}
Since $\CR(C(H'))-(r+\sigma)=\CR(C(H))-r\ge0$ and $0< r+\sigma\le(1+\beta)r$,
\[
\gamma_{r+\sigma}(H')=\frac{\CR(C(H))-r}{r+\sigma}\ge\frac{\CR(C(H))-r}{(1+\beta)r}=\frac{\gamma_r(H)}{1+\beta}.
\]
\end{IEEEproof}

Consequently, if $\sigma$ is computable in polynomial time from $(H,r)$, the mapping  $(H,r)\mapsto(H',r+\sigma)$ defines a many-one reduction between the corresponding decision problems. Since $\CR(C(H))>(1+\delta)r$ is equivalent to $\gamma_r(H)>\delta$, it also maps $\mathrm{GapCRP}_{1+\delta}(\cdot,\cdot)$ to $\mathrm{GapCRP}_{1+\delta/(1+\beta)}(\cdot,\cdot)$ whenever $\sigma\le\beta r$ on all instances considered.

\section{$\Pi_2^{\mathrm P}$-Completeness and $\Pi_2^{\mathrm P}$-Hardness of Approximation}\label{sec:pi2}

We rely on two bounded-occurrence restrictions of the $\forall\exists$-3-SAT problem to establish hardness for $\mathrm{CRP}(d_{\mathrm v},\cdot)$, $\mathrm{CRP}(\cdot,d_{\mathrm c})$, and $\mathrm{CRP}(d_{\mathrm v},d_{\mathrm c})$, and their constant-gap versions, for all fixed $d_{\mathrm v},d_{\mathrm c}\ge3$: the decision version studied by  D\"ocker et al.~\cite{DockerDornLinzSemple2020}, which we use to establish  $\Pi_2^{\mathrm P}$-completeness of the covering-radius decision problem, and the gap version studied by Haviv et al.~\cite{HavivRegevTaShma2007}, to prove $\Pi_2^{\mathrm P}$-hardness for a constant approximation gap. Both reductions use the same matrix construction, a localized variant of that of \ac{GMR}~\cite[Theorem~5.5]{GuruswamiMicciancioRegev2005}. We show that the resulting matrix $H_\Phi$ is 3-systematic, that $\CR(C(H_\Phi))\le n$ if every universal assignment can be extended to an assignment satisfying all clauses, and that $\CR(C(H_\Phi))\ge n+\lceil\nu/2\rceil$ 
if there exists a universal assignment for which every existential assignment leaves at least $\nu$ clauses unsatisfied. We then apply Theorem~\ref{thm:degree-three} with $D=3$ to $H_\Phi$, which gives a matrix whose row and column weights are at most three.

\subsection{The Source Problem $\forall\exists$-3-BSAT}\label{subsec:3bsat}

To prove $\Pi_2^{\mathrm P}$-completeness of the covering-radius decision problem, we use a bounded-occurrence restriction of the  $\forall\exists$-3-SAT problem. Let $U$ and $E$ be two disjoint sets of Boolean variables, universally and existentially quantified, respectively, and let $n=|U|+|E|$ be the total number of variables. Let $\mathcal C$ be a collection of $m$ clauses, each containing at most three literals over the variables in $U\cup E$. Without loss of generality, every clause is nonempty and contains distinct variables. A quantified Boolean formula $\Phi$ is defined as
\[
\Phi=\forall U\;\exists E\;:\bigwedge_{K\in\mathcal C}K.
\]
The $\forall\exists$-3-SAT problem asks whether, for every truth assignment to $U$, there exists a truth assignment to $E$ that satisfies at least $k$ clauses of $\mathcal C$, where the integer $k\le m$ is part of the input. We identify a truth assignment with the set of variables it sets to true; for instance, $A\cup X$ denotes the assignment in which exactly the variables of $A\cup X$ are true.

The bounded-occurrence restriction considered here, denoted by $\forall\exists$-3-BSAT, is defined by the following conditions:
\begin{enumerate}
    \item[(C1)] every universal variable occurs exactly once positively and once negatively;
    \item[(C2)] every existential variable occurs positively and negatively, and at most twice in each sign;
    \item[(C3)] $k = m$ (perfect-completeness property). 
\end{enumerate}

For an instance of $\forall\exists$-3-BSAT, we say that $\Phi$ is \emph{true} if the answer is YES, and \emph{false} otherwise.
D\"ocker et al.~\cite[Theorem~3.5]{DockerDornLinzSemple2020} prove $\Pi_2^{\mathrm P}$-completeness of the special case in which every clause has exactly three literals, $|U|=|E|$, and condition~(C2) holds in its strict form, i.e., every existential variable occurs exactly twice positively and exactly twice negatively. Every such instance is also an instance of $\forall\exists$-3-BSAT; therefore:
\begin{lemma}\label{lem:docker}
$\forall\exists$-3-BSAT is $\Pi_2^{\mathrm P}$-complete.
\end{lemma}

For $v\in U\cup E$, define
\[
\mathcal C[+v]=\{K\in\mathcal C: \text{the literal }v  \text{ occurs in }K\},
\]
\[
\mathcal C[-v]=\{K\in\mathcal C: \text{the literal }\neg v\text{ occurs in }K\}.
\]

Clearly, $\mathcal C[+v]\cap\mathcal C[-v]=\empt$ for every $v\in U\cup E$; by (C1), $|\mathcal C[+v]| = |\mathcal C[-v]| = 1$ for every $v\in U$, and by (C2), $|\mathcal C[+v]| \le 2$ and $|\mathcal C[-v]| \le 2$ for every $v\in E$.

\subsection{A Localized Column-Weight-Three Construction}\label{subsec:localized}

Let $\Phi$ be a quantified Boolean formula satisfying  assumptions (C1)--(C2).  Let $H_\Phi$ be the matrix over $\F_2$ whose rows are partitioned into a clause block
(indexed by $\mathcal C$), an ordinary-variable block (indexed by $U\cup E$), and an
additional-universal block (indexed by $U$); the row of the last block corresponding to
$x\in U$ is denoted $x^\star$. 
For $L\subseteq\mathcal C$, $S\subseteq U\cup E$, and
$A\subseteq U$, we write $[L,S,A]$ for the vector
\[
({\bf 1}_L,{\bf 1}_S,{\bf 1}_A)\in\F_2^{m+n+|U|},
\]
the concatenation of the incidence vectors of $L$, $S$ and $A$. The columns of $H_\Phi$ are:
\begin{enumerate}
\item $[L,\{x\},\{x\}]$ and $[L,\empt,\{x\}]$ for every $x\in U$ and every $L\subseteq\mathcal C[+x]$;
\item $[L,\{x\},\empt]$ for every $x\in U$ and every $L\subseteq\mathcal C[-x]$;
\item $[L,\{y\},\empt]$ for every $y\in E$ and every $L\subseteq\mathcal C[+y]$;
\item $[L,\{y\},\empt]$ for every $y\in E$ and every nonempty $L\subseteq\mathcal C[-y]$, and 
\item the clause-supported column $[L,\empt,\empt]$ for every $v\in U\cup E$ and every nonempty $L$ with $L\subseteq\mathcal C[+v]$ or $L\subseteq\mathcal C[-v]$.
\end{enumerate}

Repeated columns can occur only in Family~5; we retain one copy of each.

\begin{example}\label{ex:matrixHphi}
Let
\[
\Phi=\forall x_1\,\forall x_2\,\exists y_1\,\exists y_2\;:
(x_1\vee y_1\vee y_2)\wedge(\neg x_1\vee\neg y_1\vee\neg y_2)\wedge
(x_2\vee y_1\vee\neg y_2)\wedge(\neg x_2\vee\neg y_1\vee y_2).
\]
We have $U = \{x_1, x_2\}$, $E = \{y_1, y_2\}$, $\mathcal C[+x_1] = \{K_1\}$, $\mathcal C[-x_1] = \{K_2\}$, $\mathcal C[+x_2] = \{K_3\}$, $\mathcal C[-x_2] = \{K_4\}$, $\mathcal C[+y_1] = \{K_1, K_3\}$, $\mathcal C[-y_1] = \{K_2, K_4\}$, $\mathcal C[+y_2] = \{K_1, K_4\}$, and $\mathcal C[-y_2] = \{K_2, K_3\}$. The matrix $H_\Phi$ is therefore
\[
\setlength{\arraycolsep}{2pt}
\renewcommand{\arraystretch}{0.75}
H_\Phi=
\begin{array}{@{}r@{\;}c@{}}
 & \hphantom{\Bigg[}\begin{array}{*{34}{wc{4pt}}}
 \multicolumn{8}{c}{\scriptstyle\text{Fam.~1}} & \multicolumn{4}{c}{\scriptstyle\text{Fam.~2}} & \multicolumn{8}{c}{\scriptstyle \text{Fam.~3}} & \multicolumn{6}{c}{\scriptstyle\text{Fam.~4}} & \multicolumn{8}{c}{\scriptstyle\text{Fam.~5}}\\[-6pt]
 &&&&&&&&&&&&&&&&&&&&&&&&&&&&&&&&&
 \end{array}\hphantom{\Bigg]}\\[2pt]
\begin{array}{r}
K_1\\ K_2\\ K_3\\ K_4\\ x_1\\ x_2\\ y_1\\ y_2\\ x^\star_1\\ x^\star_2
\end{array}
&
\left[
\begin{array}{*{8}{wc{4pt}}|*{4}{wc{4pt}}|*{8}{wc{4pt}}|*{6}{wc{4pt}}|*{8}{wc{4pt}}}
 & &1&1& & & & & & & & & &1& &1& &1& &1& & & & & & &1& & & &1&1& & \\
 & & & & & & & & &1& & & & & & & & & & &1&1& &1&1& & &1& & & & &1&1\\
 & & & & & &1&1& & & & & & &1&1& & & & & & & & &1&1& & &1& &1& & &1\\
 & & & & & & & & & & &1& & & & & & &1&1& &1&1& & & & & & &1& &1&1& \\
\hline
1& &1& & & & & &1&1& & & & & & & & & & & & & & & & & & & & & & & & \\
 & & & &1& &1& & & &1&1& & & & & & & & & & & & & & & & & & & & & & \\
 & & & & & & & & & & & &1&1&1&1& & & & &1&1&1& & & & & & & & & & & \\
 & & & & & & & & & & & & & & & &1&1&1&1& & & &1&1&1& & & & & & & & \\
\hline
1&1&1&1& & & & & & & & & & & & & & & & & & & & & & & & & & & & & & \\
 & & & &1&1&1&1& & & & & & & & & & & & & & & & & & & & & & & & & & 
\end{array}
\right]
\end{array}
\]
\end{example}

The row blocks, the notation $[L,S,A]$, and the column Families~1--4 are those of the construction of \ac{GMR}~\cite[Proof of Theorem~5.5]{GuruswamiMicciancioRegev2005}; the restriction to nonempty $L$ in Family~4 only avoids repeating $[\empt,\{y\},\empt]$. Our construction differs from that of \ac{GMR} in the choice of clause-supported columns. The source problem used in~\cite{GuruswamiMicciancioRegev2005} is the gap problem Gap$\forall\exists$-3-SAT-B$_g$, shown to be $\Pi_2^{\mathrm P}$-hard by Ko and Lin~\cite{KoLin1995}: an instance comes with a threshold $t$, playing the role of our $k$, and in a YES instance every universal assignment extends to one satisfying at least $t$ clauses, not necessarily all of them. Their completeness proof must therefore also handle clauses left unsatisfied. To cover these clauses, their construction includes $[L,\empt,\empt]$ for every nonempty subset $L\subseteq\mathcal C$ of size at most the maximum number of occurrences of a variable, without requiring the clauses in $L$ to contain a common literal. In contrast, both the source problem considered here and the gap version used in Section \ref{subsec:pi2-gap} satisfy condition (C3): in a YES instance, every assignment to the universal variables can be extended to satisfy all clauses. 
Under this condition, our completeness proof needs only clause-supported columns $[L,\empt,\empt]$ whose clauses all contain a common literal (see Lemma~\ref{lem:complete}). We retain these columns in Family~5, which we call \emph{localized}. This family contributes at most $2|U|+6|E|\le6n$ columns.

As noted in~\cite[Sec.~1]{HavivRegevTaShma2007}, condition (C3) 
simplifies the proof of~\cite[Theorem~5.5]{GuruswamiMicciancioRegev2005}. In our construction, it also allows the restriction to localized clause-supported columns. The resulting matrix has at most $13n$ columns, and the weights of rows and columns are bounded by a constant (see Lemma~\ref{lem:source-properties}), as needed for the subsequent row-weight reduction.

\begin{example}[Localized clause-supported columns]\label{ex:localized-columns}
Consider the formula $\Phi$ in Example \ref{ex:matrixHphi}.
The clauses $K_1$ and $K_3$ have the literal $y_1$ in common, therefore $\{K_1,K_3\}\subseteq\mathcal C[+y_1]$. Accordingly, matrix $H_\Phi$ contains the column $[\{K_1,K_3\},\empt,\empt]$. 
The clauses $K_1$ and $K_2$, instead, share the variable $y_1$, but with opposite signs, and have no common literal. Hence, the column $[\{K_1,K_2\},\empt,\empt]$, which is included in the GMR construction, does not appear in $H_\Phi$. However, it can be expressed as the sum of the single clause-supported columns $[\{K_1\},\empt,\empt]$ and $[\{K_2\},\empt,\empt]$.
\end{example}

\begin{lemma}\label{lem:source-properties}
The matrix $H_\Phi$ (i) can be generated in polynomial time and, in particular, has at most $13n$ columns; (ii) satisfies $\Delta_{\mathrm c}(H_\Phi)\le10$; and (iii) is 3-systematic.
\end{lemma}

\begin{IEEEproof}
(i) The matrix has $m+n+|U|$ rows. Before duplicate removal, Families~1--2 contribute six columns per universal variable, Families~3--4 contribute at most seven columns per existential variable, and Family~5 contributes at most two columns per universal variable and six per existential variable. This yields at most $8|U|+13|E|\le13n$ columns, and hence $H_\Phi$ admits a polynomial-time construction.

(ii) By (C1) and (C2), rows  in the ordinary-variable block have weight at most seven, and every row in the additional-universal block has weight four.  Each clause $K$ lies in at most $2\cdot 3=6$ columns of Families~1--4, and in at most $1+3=4$ columns of Family~5 (the singleton $\{K\}$ and at most one two-clause support per existential literal of $K$). Hence $\Delta_{\mathrm c}(H_\Phi)\le 6+4=10$.

(iii) Each family contains the columns for \emph{all} admissible sets $L$ of clauses, and in particular for the smallest ones. Taking $L=\emptyset$ in Families~1--3 yields the unit vectors $[\emptyset,\emptyset,\{x\}]$, $[\emptyset,\{x\},\emptyset]$ for $x\in U$ and $[\emptyset,\{y\},\emptyset]$ for $y\in E$, while taking a singleton $L=\{K\}$ in Family~5, which is admissible because $K$ is nonempty, yields the unit vector $[\{K\},\emptyset,\emptyset]$.
Hence every unit vector is a column of $H_\Phi$, so $H_\Phi$ has full row rank. Every column is nonzero, since it has a variable entry or a nonempty $L$, and repeated columns have been removed. Moreover, every column has weight at most three: columns in Families~1--2 have at most one clause entry and at most two variable entries, columns in Families~3--4 have at most two clause entries and one variable entry, and columns in Family~5 have at most two clause entries. Hence $H_\Phi$ is 3-systematic.

\end{IEEEproof}

\subsection{Completeness and  Soundness}

The following proofs adapt the YES- and NO-case arguments of~\cite[Proof of Theorem~5.5]{GuruswamiMicciancioRegev2005} to our localized construction. Let $\Phi$ be a quantified Boolean formula with $n$ variables satisfying (C1)--(C2).
\begin{remark}\label{rem:Hphi-fullrank}
Notice that $H_\Phi$ has full row rank (Lemma~\ref{lem:source-properties}(iii)), hence $\operatorname{im}(H_\Phi)=\F_2^{m + n + |U|}$: the syndrome space of $C(H_\Phi)$ is the whole codomain, and the maximum in \eqref{eq:syndrome-radius} is taken over all of it.
\end{remark}
\begin{lemma}[Completeness]\label{lem:complete}
If $\Phi$ is a YES instance of $\forall\exists$-3-BSAT, then $\CR(C(H_\Phi))\le n$.
\end{lemma}

\begin{IEEEproof}
Let $q=[L,S,A]$ be any syndrome. Consider the assignment of the universal variables in which exactly those in $A$ are true. Since $\Phi$ is a YES instance, there exists $X\subseteq E$ such that setting to true exactly the variables in $A\cup X$ satisfies every clause. 
Let $L_v$ denote a set of clauses of $L$ containing a literal of $v$ that is true under $A\cup X$, chosen so that each clause of $L$ lies in exactly one $L_v$ (possible since $A\cup X$ satisfies every clause).  
For every variable $v$, set
\[
h_v=[L_v,\{v\}\cap S,\{v\}\cap A],
\]
and let us verify that every nonzero $h_v$ is a column of $H_\Phi$. If $v \in A$, $h_v$ belongs to Family~1. 
If $v \in (U\backslash A)\cap S$, $h_v$ belongs to Family~2. If $v$ is an existential variable in $S$, then $h_v$ belongs to Family~3 when $v$ is true or $L_v =\empt$, and to Family~4 otherwise. In all remaining cases, $h_v=[L_v,\empt,\empt]$, which either is $\bbzero$ or belongs to Family~5, since $L_v$ is contained in $\mathcal C[+v]$ or $\mathcal C[-v]$.

We thus have that the sets $L_v$, $v \in U \cup E$, are pairwise disjoint with union $L$, that each variable in $S$ contributes exactly once to the ordinary-variable block of rows, and that each universal variable in $A$ contributes exactly once to the additional-universal block of rows. Consequently, 
\[
\sum_{v\in U\cup E}h_v=[L,S,A]=q,
\]
where the sum is over $\mathbb F_2$. The claim follows from the fact that $q$ is the sum of at most $n$ columns of $H_\Phi$.
\end{IEEEproof}

\medskip

\begin{lemma}[Soundness]\label{lem:sound}
Let $A\subseteq U$ and $\nu\ge 0$ be such that, for every $X\subseteq E$, at least $\nu$ clauses are left unsatisfied by setting to true exactly the variables in $A\cup X$. Then $\bar q=[\mathcal C,U\cup E,A]$ satisfies $\lambda_{H_\Phi}(\bar q)\ge n+\lceil\nu/2\rceil$. In particular, if $\Phi$ is a NO instance of $\forall\exists$-3-BSAT, then $\CR(C(H_\Phi))\ge n + 1$. 
\end{lemma}

\begin{IEEEproof}
Let $\bar q =[\mathcal C,U\cup E,A]$. By construction, each column of $H_\Phi$ has at most one entry equal to 1 in the ordinary-variable block, so every set $\mathcal R$ of columns of $H_\Phi$ summing to $\bar q$ must contain, for each variable $v\in U\cup E$, at least one column with a $1$ in row $v$ of the ordinary-variable block (a \emph{$v$-column}); hence $|\mathcal R|\ge n$. 
For each $v$, select one $v$-column $c_v\in\mathcal R$, called the \emph{selected column} of $v$, and let $t=|\mathcal R|-n$ be the number of non-selected columns.

For $y\in E$, the selected column $c_y$ is $[L_y,\{y\},\empt]$ with $L_y\subseteq\mathcal C[+y]$ (Family~3) or $L_y\subseteq\mathcal C[-y]$ (Family~4); set $y$ to true if $\empt\ne L_y\subseteq\mathcal C[+y]$ and to false otherwise, and let $X\subseteq E$ be the set of existential variables set to true. For $x\in U$, the selected column $c_x$ is either $[L_x,\{x\},\{x\}]$ with $L_x\subseteq\mathcal C[+x]$ (Family~1) or $[L_x,\{x\},\empt]$ with $L_x\subseteq\mathcal C[-x]$ (Family~2). Call $c_x$ \emph{wrong} if it belongs to Family~2 and $x\in A$, or to Family~1 and $x\notin A$; all other selected columns, including every $c_y$ with $y\in E$, are called \emph{correct}. If $c_x$ is wrong, the parity of row $x^\star$ of $\bar q$ forces a non-selected column $p_x\in\mathcal R$ with a $1$ in row $x^\star$. It belongs to Family~1, so its clause entries lie in $\mathcal C[+x]$. The forced non-selected columns $p_x$ are pairwise distinct, since each column has at most one $1$ in the additional-universal block. 

Let $\bar{\mathcal C}$ be the set of clauses left unsatisfied by $A\cup X$ (recall that $|\bar{\mathcal C}|\ge\nu$), and let $\bar K\in\bar{\mathcal C}$. Since $\bar K$ is falsified by the truth assignment determined by $A$ and $X$, every literal of $\bar K$ is false. Hence, for each variable $v$ occurring in $\bar K$, the literal is $\neg v$ if $v\in A\cup X$, and $v$ if $v\notin A\cup X$. On the other hand, every correct column $c_v$ has its clause entries in $\mathcal C[\ell]$ for a literal $\ell$ of $v$ that is true under $A\cup X$: for $x\in U$, $\ell=x$ if $x\in A$ (Family~1) and $\ell=\neg x$ if $x\notin A$ (Family~2); for $y\in E$, $\ell=y$ if $y\in X$ and $\ell=\neg y$ otherwise, by the choice of $X$. Therefore every correct column has a $0$ in row $\bar K$. Since $\bar q$ has a $1$ there, row $\bar K$ is covered by a wrong column or by a non-selected column. Let $b$ be the number of wrong columns. By (C1), each wrong $c_x$ and each $p_x$ has at most one clause entry, and by (C1)--(C2) every other non-selected column has at most two. Hence $|\bar{\mathcal C}|\le b+b+2(t-b)=2t$, and $t\ge\lceil\nu/2\rceil$.

Therefore every such $\mathcal R$ satisfies $|\mathcal R|=n+t\ge n+\lceil\nu/2\rceil$, that is, $\lambda_{H_\Phi}(\bar q)\ge n+\lceil\nu/2\rceil$.

If $\Phi$ is a NO instance, there is $A\subseteq U$ such that every $X\subseteq E$ leaves at least one clause unsatisfied, that is, the hypothesis holds with $\nu=1$. Then $\lambda_{H_\Phi}(\bar q)\ge n+1$, and since $\bar q$ is a syndrome (Remark~\ref{rem:Hphi-fullrank}), the claim follows by the definition of covering radius.
\end{IEEEproof}

\begin{proposition}\label{prop:base-reduction}
Given a quantified Boolean formula $\Phi$ satisfying assumptions (C1)--(C2), one can construct in polynomial time a 3-systematic matrix $H_\Phi$ with $\Delta_{\mathrm v}(H_\Phi)\le3$ and $\Delta_{\mathrm c}(H_\Phi)\le10$ such that
\[
\Phi\text{ is true}
\quad\Longleftrightarrow\quad
\CR(C(H_\Phi))\le n.
\]
\end{proposition}

\begin{IEEEproof}
Lemma~\ref{lem:source-properties} gives the structural properties, and Lemmas~\ref{lem:complete} and~\ref{lem:sound} give the equivalence.
\end{IEEEproof}

\subsection{$\Pi_2^{\mathrm P}$-Completeness}

\begin{proposition}\label{prop:crp33}
The problem $\mathrm{CRP}(3,3)$ is $\Pi_2^{\mathrm P}$-complete.
\end{proposition}

\begin{IEEEproof}
Membership in $\Pi_2^{\mathrm P}$ follows from Section~\ref{sec:preli}. For $\Pi_2^{\mathrm P}$-hardness,  a many-one reduction from $\forall\exists$-3-BSAT, which is $\Pi_2^{\mathrm P}$-complete by Lemma~\ref{lem:docker}, is as follows. Given an instance $\Phi\in \forall\exists\text{-3-BSAT}$, construct $H_\Phi$ as described in  Proposition~\ref{prop:base-reduction}. Let $H_\Phi^{(3)}$ be the matrix obtained from $H_\Phi$ by Theorem~\ref{thm:degree-three} with $D=3$. By Proposition~\ref{prop:base-reduction} and Theorem~\ref{thm:degree-three},
\[
\Delta_{\mathrm v}(H_\Phi^{(3)})\le3,
\qquad
\Delta_{\mathrm c}(H_\Phi^{(3)})\le3,
\]
and
\begin{align*}
\Phi\text{ true}
&\iff \CR(C(H_\Phi))\le n\\
&\iff \CR(C(H_\Phi^{(3)}))\le n+\Omega_3(H_\Phi).
\end{align*}
Both constructions are polynomial-time, and $\Omega_3(H_\Phi)$ is computable directly from $H_\Phi$. 
\end{IEEEproof}

\begin{restatement}{Theorem~\ref{thm:bounded}(ii)}
For all fixed $d_{\mathrm v},d_{\mathrm c}\ge3$, $\mathrm{CRP}(d_{\mathrm v},d_{\mathrm c})$, $\mathrm{CRP}(\cdot,d_{\mathrm c})$ and $\mathrm{CRP}(d_{\mathrm v},\cdot)$ are $\PiP$-complete.
\end{restatement}

\begin{IEEEproof}
The $\Pi_2^{\mathrm P}$-hardness follows from Proposition~\ref{prop:crp33} and from the fact that $\mathrm{CRP}(3,3)$ is a restriction of $\mathrm{CRP}(d_{\mathrm v},d_{\mathrm c})$, and $\mathrm{CRP}(d_{\mathrm v},d_{\mathrm c})$ is a restriction of both $\mathrm{CRP}(\cdot,d_{\mathrm c})$ and $\mathrm{CRP}(d_{\mathrm v},\cdot)$ for any fixed $d_{\mathrm v},d_{\mathrm c}\ge3$. 
Membership in $\Pi_2^{\mathrm P}$ follows from Section~\ref{sec:preli}.
\end{IEEEproof}

\subsection{$\Pi_2^{\mathrm P}$-Hard Constant Gaps}\label{subsec:pi2-gap}

We next prove $\Pi_2^{\mathrm P}$-hardness for a constant approximation gap. 
Since the decision version of $\forall\exists$-3-SAT gives no guarantee on the number of clauses that remain unsatisfied in a NO instance, hardness-of-approximation results require a promise version $\mathrm{Gap}_\varepsilon \forall\exists$-3-SAT, for $0 < \varepsilon\le 1$, of the problem, see Haviv et al.~\cite{HavivRegevTaShma2007}: given a quantified Boolean formula $\Phi$, it is a YES instance of $\mathrm{Gap}_\varepsilon \forall\exists$-3-SAT if, for every assignment to the universal variables, there exists an assignment to the existential variables satisfying all clauses. $\Phi$ is a NO instance if there exists an assignment to the universal variables such that every assignment to the existential variables leaves at least an $\varepsilon$ fraction of the clauses unsatisfied.

\begin{lemma}\label{lem:hardGapSAT}
There exists a constant $0 < \varepsilon\le 1$ for which the gap version $\mathrm{Gap}_\varepsilon\forall\exists$-3-BSAT of $\forall\exists$-3-BSAT, i.e., $\mathrm{Gap}_\varepsilon\forall\exists$-3-SAT restricted to formulas satisfying (C1)--(C2), is $\Pi_2^{\mathrm P}$-hard.
\end{lemma}

\begin{IEEEproof}
The polynomial reduction is from the $\mathrm{Gap}_\varepsilon \forall\exists$-3-SAT problem in which each universal variable occurs at most twice and each existential variable occurs at most three times, a restriction under which the problem is still $\Pi_2^{\mathrm P}$-hard for a constant $0 < \varepsilon\le 1$ (see Haviv, Regev, and Ta-Shma~\cite[Theorem~1.1, p.~46, and Sec.~4]{HavivRegevTaShma2007}).

Let $\Phi$ be an instance of the above problem. Tautological clauses and duplicate occurrences of the same literal within a clause can be removed. Moreover, pure variables can be repeatedly eliminated (see~\cite[Proof of Theorem~5.5]{GuruswamiMicciancioRegev2005}) as follows: each pure existential variable is assigned the truth value that satisfies all its occurrences, after which the clauses containing it are removed; each pure universal variable is instead assigned the truth value that falsifies all its occurrences, after which these literals are removed from their clauses. Variables that no longer occur in any clause are removed from the corresponding variable set.

After pure-variable elimination, every remaining universal variable occurs exactly once in each sign, and every existential variable occurs in both signs and at most three times in total, so each signed occurrence set has size at most two. Every clause contains at most three literals on distinct variables. Unless a clause becomes empty or all clauses are deleted, the resulting formula $\Phi'$ is therefore an instance of $\mathrm{Gap}_\varepsilon\forall\exists$-3-BSAT.

These operations map YES instances $\Phi$ to YES instances $\Phi'$ and preserve the NO-case gap with the same $\varepsilon$: the removed clauses are satisfied, so the number of unsatisfied clauses does not change while the number of clauses can only decrease, and the falsifying value of a pure universal variable is the best choice for the universal player. If a clause becomes empty, $\Phi$ cannot be a YES instance, and we set $\Phi'=\exists y:(y)\wedge(\neg y)$. If all clauses are deleted, we set $\Phi'=\exists y_1\exists y_2:(y_1\vee y_2)\wedge(\neg y_1\vee\neg y_2)$. Both satisfy (C1)--(C2), and the first leaves one of its two clauses unsatisfied under every assignment, so it is a NO instance whenever $\varepsilon\le1/2$. Hence the lemma holds with the constant $\min\{\varepsilon,1/2\}$.
\end{IEEEproof}

Let $\Phi$ be an instance of $\mathrm{Gap}_\varepsilon\forall\exists$-3-BSAT, with $n$ variables and $m$ clauses. By (C1)--(C2), every variable occurs at least twice, and every clause has at most three literals, so
\begin{equation}\label{eq:gap-mn}
\frac{2n}{3}\le m.
\end{equation}

Lemma~\ref{lem:source-properties} gives a 3-systematic matrix $H_\Phi$ with $\Delta_{\mathrm v}(H_\Phi)\le3$, $\Delta_{\mathrm c}(H_\Phi)\le10$, and at most $13n$ columns. Since every column has weight at most three,
\begin{equation}\label{eq:gap-nnz}
\nnz(H_\Phi)\le39n.
\end{equation}

For a YES instance, Lemma~\ref{lem:complete} gives $\CR(C(H_\Phi))\le n$. For a NO instance, let $A$ be the universal assignment guaranteed by the promise. Every existential assignment leaves at least $\varepsilon m$ clauses unsatisfied. Lemma~\ref{lem:sound} and~\eqref{eq:gap-mn} therefore give
\begin{equation}\label{eq:gap-base-no}
\CR(C(H_\Phi))
\ge n+\frac{\varepsilon m}{2}
\ge \left(1+\frac{\varepsilon}{3}\right)n.
\end{equation}

Let $H_\Phi^{(3)}$ be the matrix obtained from $H_\Phi$ by Theorem~\ref{thm:degree-three} with $D=3$, and let $r_3=n+\Omega_3(H_\Phi)$.
Since $\Omega_3(H_\Phi)$ is the total weight of the rows of $H_\Phi$ of weight at least four,~\eqref{eq:gap-nnz} gives $r_3-n=\Omega_3(H_\Phi)\le\nnz(H_\Phi)\le39n$. By Theorem~\ref{thm:degree-three}, $\CR(C(H_\Phi^{(3)}))=\CR(C(H_\Phi))+\Omega_3(H_\Phi)$, so a YES instance of $\mathrm{Gap}_\varepsilon\forall\exists$-3-BSAT gives $\CR(C(H_\Phi^{(3)}))\le r_3$.
For a NO instance, we can apply Lemma~\ref{lem:gap-transfer} with $r=n$, $\delta=\varepsilon/3$ and $\beta=39$, since $\gamma_n(H_\Phi) \ge \varepsilon/3$ by~\eqref{eq:gap-base-no} and $r_3-n\le39n$. Consequently, the lemma yields
\[
\gamma_{r_3}(H_\Phi^{(3)}) \ge \frac{\varepsilon}{120},
\]
that is, $\CR(C(H_\Phi^{(3)}))\ge(1+\varepsilon/120)\,r_3>(1+\varepsilon/121)\,r_3$, where the strict inequality uses $r_3\ge n>0$. The strict NO condition in the definition of $\mathrm{GapCRP}_{1+c_3}$ is therefore satisfied by choosing, e.g., $c_3=\varepsilon/121$. 

\begin{restatement}{Theorem~\ref{thm:bounded}(iii)}
There is a constant $c>0$ such that $\mathrm{GapCRP}_{1+c}(3,3)$ is $\PiP$-hard.
\end{restatement}

\begin{IEEEproof}
The construction above is a polynomial-time reduction from $\mathrm{Gap}_\varepsilon\forall\exists$-3-BSAT, which is $\PiP$-hard by Lemma~\ref{lem:hardGapSAT}, to $\mathrm{GapCRP}_{1+c_3}(3,3)$, with $c_3=\varepsilon/121>0$.
\end{IEEEproof}

Since every instance of $\mathrm{GapCRP}_{1+c_3}(3,3)$ is also an instance of $\mathrm{GapCRP}_{1+c_3}(d_{\mathrm v},d_{\mathrm c})$, $\mathrm{GapCRP}_{1+c_3}(\cdot,d_{\mathrm c})$ and $\mathrm{GapCRP}_{1+c_3}(d_{\mathrm v},\cdot)$ for all $d_{\mathrm v},d_{\mathrm c}\ge3$, these problems are $\PiP$-hard as well.

\section{Explicit Constants for NP-Hardness of Approximation}\label{sec:np}

Section~\ref{sec:pi2} establishes the existence of a constant
$\alpha>1$ for which approximating the covering radius within a factor
$\alpha$ is $\Pi_2^{\mathrm P}$-hard, but it does not give an explicit numerical value of~$\alpha$, since $\alpha=1+\varepsilon/121$ depends on the constant $\varepsilon$ of Lemma~\ref{lem:hardGapSAT}. 
Explicit factors, however, can be obtained for NP-hardness of approximation. To this end, we now reduce from Minimum Vertex Cover on 3-regular graphs, which is NP-hard to approximate within an explicit constant factor~\cite[Corollary~18]{ChlebikChlebikova2006}. 
Applying the Set-Cover construction of \ac{GMR}~\cite[Proof of Theorem~5.7]{GuruswamiMicciancioRegev2005} to a 3-regular Vertex Cover instance yields the 3-systematic matrix constructed below, whose covering radius equals the size of a minimum vertex cover. The transformations of Section~\ref{sec:row-reduction} then lower the row weights to the required bound, keeping the column weights at most three, while preserving a constant multiplicative gap.

Before deriving these explicit factors, we show that the rank of a parity-check matrix provides a polynomial-time approximation for the covering radius problem with bounded column weight or bounded row weight. For general binary linear codes, \ac{GMR}~\cite[Theorem~5.2]{GuruswamiMicciancioRegev2005} show that $\operatorname{rank}(H)$ approximates $\CR(C(H))$ within a factor logarithmic in $N$. The following proposition shows that, if $\Delta_{\mathrm v}(H)\le d_{\mathrm v}$, the rank lies between the covering radius and $d_{\mathrm v}$ times the covering radius, giving an approximation factor of $d_{\mathrm v}$.

\begin{proposition}\label{prop:dv-approx}
Let $d_{\mathrm v}\ge2$ and let $H\in\F_2^{s\times N}$ satisfy $\Delta_{\mathrm v}(H)\le d_{\mathrm v}$. Then
\begin{equation}\label{eq:dv-approx}
\left\lceil\frac{\operatorname{rank}(H)}{d_{\mathrm v}}\right\rceil
\le \CR(C(H))\le \operatorname{rank}(H).
\end{equation}
Consequently, $\mathrm{GapCRP}_{d_{\mathrm v}}(d_{\mathrm v},\cdot)$ is solvable in polynomial time. 
\end{proposition}

\begin{IEEEproof}
If $\operatorname{rank}(H)=0$, then $C(H)=\F_2^N$ and all three quantities in~\eqref{eq:dv-approx} vanish. Otherwise, by Lemma~\ref{lem:normalization}, we may assume that $H$ has full row rank, and hence exactly $\operatorname{rank}(H)$ rows. The upper bound is the redundancy bound~\cite[Corollary~11.1.3]{HuffmanPless2003}. For the lower bound, the syndrome $\bbone$ is attainable (because of full rank) and has weight $\operatorname{rank}(H)$, while a sum of $t$ columns of weight at most $d_{\mathrm v}$ has weight at most $t\,d_{\mathrm v}$. Hence every representative of the syndrome $\bbone$ has weight at least $\lceil\operatorname{rank}(H)/d_{\mathrm v}\rceil$.  Therefore, a YES instance $(H,r)$ of $\mathrm{GapCRP}_{d_{\mathrm v}}(d_{\mathrm v},\cdot)$ satisfies, by~\eqref{eq:dv-approx}, $\operatorname{rank}(H)\le d_{\mathrm v}\CR(C(H))\le d_{\mathrm v}r$. On the other hand, a NO instance satisfies $\operatorname{rank}(H)\ge\CR(C(H))>d_{\mathrm v}r$. Hence comparing $\operatorname{rank}(H)$ with $d_{\mathrm v}r$ distinguishes the two cases, and since the rank is computable in polynomial time, the problem is in $\mathrm{P}$.

\end{IEEEproof}

For matrices with row and column weights at most three, Proposition~\ref{prop:dv-approx} gives a polynomial-time approximation factor of three. 

A bound on the row weight alone also gives a constant factor.  The argument below refines the volume bound behind~\cite[Theorem~5.2]{GuruswamiMicciancioRegev2005}.   Let $h_2(p)=-p\log_2p-(1-p)\log_2(1-p)$ denote the binary entropy function.

\begin{proposition}\label{prop:dc-approx}
Let $d_{\mathrm c}\ge3$, let $p_{d_{\mathrm c}}\in(0,1/2)$ be such that $h_2(p_{d_{\mathrm c}})=1/d_{\mathrm c}$, and let $H\in\F_2^{s\times N}$ satisfy $\Delta_{\mathrm c}(H)\le d_{\mathrm c}$. Then
\begin{equation}\label{eq:dc-approx}
d_{\mathrm c}\,p_{d_{\mathrm c}}\operatorname{rank}(H)\le\CR(C(H))\le\operatorname{rank}(H).
\end{equation}
Consequently, $\mathrm{GapCRP}_{\alpha_{d_{\mathrm c}}}(\cdot,d_{\mathrm c})$ is solvable in polynomial time for $\alpha_{d_{\mathrm c}}=1/(d_{\mathrm c}\,p_{d_{\mathrm c}})$. In particular, $\alpha_3<5.43$.
\end{proposition}

\begin{IEEEproof}
The upper bound is the redundancy bound, as in Proposition~\ref{prop:dv-approx}. For the lower bound, the case $\operatorname{rank}(H)=0$ is trivial. Otherwise, by Lemma~\ref{lem:normalization}, we may assume that $H$ has full row rank and no $\bbzero$-columns. Hence, since $H$ has $\operatorname{rank}(H)$ rows of weight at most $d_{\mathrm c}$, $N\le\nnz(H)\le d_{\mathrm c}\operatorname{rank}(H)$. Let $V(M,t)=\sum_{i=0}^{t}\binom{M}{i}$ be the number of vectors in a Hamming ball of radius $t$ in $\F_2^M$.  The $2^{N-\operatorname{rank}(H)}$ balls of radius $\CR(C(H))$ centered at the codewords cover $\F_2^N$, so $2^{N-\operatorname{rank}(H)}V(N,\CR(C(H)))\ge2^N$ (sphere-covering bound~\cite{CohenHonkalaLitsynLobstein1997}). Since $V(M,t)$ is nondecreasing in $M$ and $N\le d_{\mathrm c}\operatorname{rank}(H)$,
\[
2^{\operatorname{rank}(H)}\le V\bigl(N,\CR(C(H))\bigr)\le V\bigl(d_{\mathrm c}\operatorname{rank}(H),\CR(C(H))\bigr).
\]
For $p=t/M\le1/2$, the terms $p^i(1-p)^{M-i}$ do not increase with $i$, so
\[
1=\sum_{i=0}^{M}\binom{M}{i}p^i(1-p)^{M-i}\ge V(M,t)\,p^{t}(1-p)^{M-t}=V(M,t)\,2^{-M h_2(p)},
\]
that is, $V(M,t)\le2^{M h_2(t/M)}$. Applying this with $M=d_{\mathrm c}\operatorname{rank}(H)$ and $t=\CR(C(H))$, which is allowed since $\CR(C(H))\le\operatorname{rank}(H)$, gives
\[
h_2\left(\frac{\CR(C(H))}{d_{\mathrm c}\operatorname{rank}(H)}\right)\ge\frac{1}{d_{\mathrm c}}.
\]
Since $h_2$ is increasing on $[0,1/2]$, $\CR(C(H))\ge d_{\mathrm c}\,p_{d_{\mathrm c}}\operatorname{rank}(H)$, which is the lower bound in~\eqref{eq:dc-approx}. Finally, a YES instance $(H,r)$ satisfies $\operatorname{rank}(H)\le\CR(C(H))/(d_{\mathrm c}p_{d_{\mathrm c}})\le\alpha_{d_{\mathrm c}}r$, and a NO instance satisfies $\operatorname{rank}(H)\ge\CR(C(H))>\alpha_{d_{\mathrm c}}r$, so comparing $\operatorname{rank}(H)$ with $\alpha_{d_{\mathrm c}}r$ distinguishes the two cases in polynomial time. In particular, $h_2(0.0614)<1/3$ gives $p_3>0.0614$, hence $\alpha_3<1/(3\cdot0.0614)<5.43$.
\end{IEEEproof}

The rest of this section shows that approximating the covering radius within the factor $1882/1881$ of Theorem~\ref{thm:explicit} is NP-hard.

\subsection{Covering Radius as Minimum Vertex Cover on 3-Regular Graphs}

Let $F =(V, \mathcal E)$ be a simple $3$-regular graph, and let $\tau(F)$ denote the size of a minimum vertex cover of $F$. For each vertex $v\in V$, let $\partial(v) \subseteq\mathcal E$ be the star of $v$, i.e., the set of the three edges incident to $v$. Following the construction of \ac{GMR}~\cite[Proof of Theorem~5.7]{GuruswamiMicciancioRegev2005}, let $H$ be the binary matrix whose rows are indexed by $\mathcal E$ and which has one column ${\bf 1}_B\in\mathbb{F}_2^{\mathcal E}$  for each vertex $v\in V$ and each nonempty subset $B\subseteq\partial(v)$.  
After retaining only one copy of each repeated column, the matrix $H$ has the following structural properties:

\begin{proposition}\label{prop:cubic-structure}
The matrix $H$ is 3-systematic, every row has weight seven, and it has $\frac{11}{3}|{\mathcal E}|$ columns: $|{\mathcal E}|$ of weight one, $2|{\mathcal E}|$ of weight two, and $\frac{2}{3}|{\mathcal E}|$ of weight three.
\end{proposition}

\begin{IEEEproof}
By construction, the columns of $H$ are nonzero, pairwise distinct, and of weight at most three. Moreover, the unit column ${\bf 1}_{\{f\}}$ occurs for every edge $f$, so $H$ is 3-systematic (Definition~\ref{def:admissible}) and hence has full row rank.

An edge $f=uv$ belongs to four subsets of $\partial(u)$ and to four subsets of $\partial(v)$. Since $F$ is simple, the singleton $\{f\}$ is the only subset common to both $\partial(u)$ and $\partial(v)$. After removing repeated columns, every row therefore has weight $4+4-1=7$. Since $|V|=\frac{2}{3}|{\mathcal E}|$, counting the singleton, two-edge, and three-edge subsets gives the stated column counts. 
\end{IEEEproof}
The construction of \ac{GMR} gives the following identity.

\begin{lemma}\label{lem:cubic-gmr}
For the matrix $H$ constructed above,
\begin{equation}\label{eq:cubic-exact}
\CR(C(H))=\tau(F).
\end{equation}
\end{lemma}

\begin{IEEEproof}
By~\cite[Proof of Theorem~5.7]{GuruswamiMicciancioRegev2005}, the covering radius equals the minimum number of subsets required to cover the ground set, that is, in our case, the minimum number of stars $\partial(v)$ needed to cover all edges, which is $\tau(F)$. Deleting zero and repeated columns preserves the covering radius by Lemma~\ref{lem:copies}.
\end{IEEEproof}

For 3-regular graphs, the following inapproximability result holds:

\begin{lemma}[{Chleb\'ik and Chleb\'ikov\'a~\cite{ChlebikChlebikova2006}}]\label{lem:cc}
Given a $3$-regular graph $F= (V, \mathcal E)$ and an integer $\kappa$, it is NP-hard to distinguish $\tau(F)\le\kappa$ from $\tau(F)>\frac{100}{99}\kappa$.
\end{lemma}
\begin{IEEEproof}
For maximum degree three, the proof of~\cite[Theorem~17]{ChlebikChlebikova2006},  together with~\cite[Theorem~16 and Corollary~18]{ChlebikChlebikova2006}, gives constants $0<a<b$ with $b/a>100/99$ such that it is NP-hard to decide, for a $3$-regular graph $F= (V, \mathcal E)$, whether $\tau(F)<a|V|$ or $\tau(F)>b|V|$; in fact, any ratio $b/a<1.0101215$ is attainable. With $\kappa=\lfloor a|V|\rfloor$, the first case gives $\tau(F)\le\kappa$ and the second gives $\tau(F)>\frac{100}{99}a|V|\ge\frac{100}{99}\kappa$.
\end{IEEEproof}

By~\eqref{eq:cubic-exact}, the map $(F,\kappa)\mapsto(H,\kappa)$ already shows that $\mathrm{GapCRP}_{100/99}(3,7)$ is NP-hard on 3-systematic matrices with
all row weights seven (Proposition~\ref{prop:cubic-structure}). The subsequent reduction lowers the row weights while preserving a constant multiplicative gap.

We may assume $\kappa\ge |\mathcal E|/3$: since each vertex covers at most three edges, $|\mathcal E|\le3\tau(F)$, so under the promise any instance with $\kappa<|\mathcal E|/3$ is necessarily a NO instance and can be mapped directly to a fixed promised NO instance satisfying the required matrix constraints, namely $([1],0)$.

Theorem~\ref{thm:degree-three} with $D=3$ would bring the row weights of $H$ down to three with a shift of $\Omega_3(H)=7|\mathcal E|$. Using \textsc{CCRSplit} instead of \textsc{CRSplit} lowers the shift of each split by one (Lemma~\ref{lem:contraction}), and hence improves the approximation factor.

\subsection{Bounded $(3,3)$ Matrices}\label{subsec:np-33}

Let $H''$ be obtained by applying \textsc{CCRSplit} to each of the $|\mathcal E|$ rows of $H$, with the partition of its support into seven singletons. Every row $R_j$ of $H''$ has weight $1+2=3$, and every row $W_j$ has weight two. The columns inherited from $H$ keep their weight, which is at most three, and the new columns have weight two. Hence $\Delta_{\mathrm v}(H'')\le3$ and $\Delta_{\mathrm c}(H'')\le3$. Since $H$ is 3-systematic (Proposition~\ref{prop:cubic-structure}), Lemma~\ref{lem:contraction} shows that $H''$ has pairwise distinct nonzero columns, and that each split increases the covering radius by $7-1=6$. By~\eqref{eq:cubic-exact},
\begin{equation}\label{eq:cubic-33-radius}
\CR(C(H''))=\tau(F)+6|\mathcal E|.
\end{equation}

Set
\[
\widehat r=\kappa+6|\mathcal E|.
\]
In the YES case,~\eqref{eq:cubic-33-radius} gives $\CR(C(H''))\le\widehat r$. In the NO case,
\[
\CR(C(H''))>\widehat r+\frac{\kappa}{99}.
\]
Since $|\mathcal E|\le3\kappa$, the shift satisfies $6|\mathcal E|\le18\kappa$, and Lemma~\ref{lem:gap-transfer} with $\delta=1/99$ and $\beta=18$ gives
\[
\CR(C(H''))>
\left(1+\frac1{1881}\right)\widehat r.
\]
Hence, the following holds.

\begin{restatement}{Theorem~\ref{thm:explicit}}
$\mathrm{GapCRP}_{1882/1881}(3,3)$ is NP-hard.
\end{restatement}

\begin{IEEEproof}
The construction above, with the instances with $\kappa<|\mathcal E|/3$ mapped to $([1],0)$, is a polynomial-time reduction from the promise problem of Lemma~\ref{lem:cc} to $\mathrm{GapCRP}_{1882/1881}(3,3)$.
\end{IEEEproof}

A larger row-weight bound $D$ requires fewer splits: for $D=4,5,6$, \textsc{CCRSplit} brings each row of $H$ to weight at most $D$ with a shift of $3$, $2$ and $1$, respectively, instead of $6$. Repeating the argument above shows that $\mathrm{GapCRP}_{\alpha}(3,D)$ is NP-hard with $\alpha=991/990$, $694/693$ and $397/396$ for $D=4,5,6$; for $D\ge7$ no split is needed, and $\alpha=100/99$. We state Theorem~\ref{thm:explicit} for $(3,3)$, the smallest pair of bounds for which hardness holds.

\section{Conclusion}\label{sec:concl}

The main message of this work is that local sparsity does not remove the intrinsic difficulty of covering radius, either for exact computation or for approximation, at either of the first two levels of the polynomial hierarchy. Maximum column and row weight three already make the exact problem $\PiP$-complete and a constant multiplicative gap $\PiP$-hard, as for unrestricted linear codes~\cite{McLoughlin1984,GuruswamiMicciancioRegev2005}, whereas maximum column or row weight two makes the problem polynomial.

The main open question concerns approximation. For matrices with row and column weights at most three, approximating the covering radius within $1882/1881$ is NP-hard (Theorem~\ref{thm:explicit}), while comparing the rank with the threshold approximates it within a factor of three in polynomial time (Proposition~\ref{prop:dv-approx}). What is the smallest factor achievable in polynomial time on this class? Answering this requires either an algorithm that beats the rank bound or a hardness proof for a larger constant. More generally, for column weight at most $d_{\mathrm v}$, it is open whether any factor smaller than $d_{\mathrm v}$ can be achieved, and for row weight at most three the range between $1882/1881$ and $\alpha_3<5.43$ (Proposition~\ref{prop:dc-approx}) is open as well. This differs from general linear codes, for which every constant factor is NP-hard~\cite{GuruswamiMicciancioRegev2005}: a bound on the column weight (or on the row weight) alone already changes the approximability of the problem. 

\bibliographystyle{IEEEtran}
\bibliography{ref}

\section*{AI Disclosure Statement}
The authors used Claude Opus 5.5 (Anthropic) only to edit the language and organization of the manuscript. The authors themselves produced all statements and proofs and take full responsibility for the content.

\end{document}